\documentclass[10pt,journal]{IEEEtran}
\IEEEoverridecommandlockouts
\pdfoutput=1
\usepackage{cite}
\usepackage{amsmath,amssymb,amsfonts}
\usepackage{algorithmic}
\usepackage{graphicx}
\usepackage{subfigure}
\usepackage{textcomp}
\usepackage[table]{xcolor}
\usepackage{booktabs}
\usepackage{epstopdf}
\usepackage{multirow}   %使用multirow必须加载该package
\usepackage{soul}
\usepackage{enumerate}
 \usepackage{amssymb}
\soulregister\cite7
\soulregister\ref7

\begin{document}

\title{%Network-Load Estimation for $ K $-Repetition Grant-Free Access Enabling Adaptive Resource Allocation Towards QoS Enhancement
	Improved Grant-Free Access for URLLC via Multi-Tier-Driven Computing: Network-Load Learning, Prediction, and Resource Allocation\\}

\author{Zixiao Zhao,~\IEEEmembership{Graduate Student Member,~IEEE,} Qinghe Du,~\IEEEmembership{Member,~IEEE,} and George K. Karagiannidis,~\IEEEmembership{Fellow,~IEEE}
\thanks{The reported research work in this paper is supported by National Key Research and Development Program of China (2020YFB1807700).}
\thanks{Z. Zhao and Q. Du are with School of Information and Communications Engineering, Xi'an Jiaotong University, China and Shaanxi Smart Networks and Ubiquitous Access Research Center, China. (e-mail:zzx67120787@stu.xjtu.edu.cn, duqinghe@mail.xjtu.edu.cn). Correspondence author: Qinghe Du.}
\thanks{
G. K. Karagiannidis is with Department of Electrical and Computer En- gineering, Aristotle University of Thessaloniki, Thessaloniki 54 124, Greece. (e-mail: geokarag@auth.gr).}
\thanks{Part of the research work reported in this paper was presented in IEEE PIMRC 2021.}}

%\author{
%	\IEEEauthorblockN{Zixiao Zhao$^{1,3} $, Qinghe Du$^{1,3} $ and Li Sun$^{1,2} $}	
%	\IEEEauthorblockA{1 School of Information and Communications Engineering, Xi'an Jiaotong University, China}
%	 \IEEEauthorblockA{2 Zhejiang Laboratory, Hangzhou, 311121, China}
%	\IEEEauthorblockA{3 National Simulation Education Center for Communications and Information Systems, Xi'an Jiaotong University, China} 
%	\IEEEauthorblockA{Emails: $ \left\{\textit{zzx67120787@stu.xjtu.edu.cn, \{duqinghe, lisun\}@mail.xjtu.edu.cn}\right\} $}
%}
%
%\author{
%	\IEEEauthorblockN{\normalsize Zixiao Zhao$^{1,2} $ and Qinghe Du$^{1,2} $
%		\\
%1 School of Information and Communications Engineering, Xi'an Jiaotong University, China
%\\
%2 Shaanxi Smart Networks and Ubiquitous Access Research Center, China
%\\
%Emails: \{\textit{zzx67120787@stu.xjtu.edu.cn, duqinghe@mail.xjtu.edu.cn}\} 
%}}
\maketitle

\begin{abstract}
%In the next generation communication systems, Ultra-Reliable and Low-Latency Communications (URLLC) will support more sophisticated vertical services by fully satisfying various requirements. $K$-repetition Grant-Free (GF) access can effectively lower the latency by avoiding the hand-shake procedures and thus has the potential to provide better QoS. However, it can hardly achieve both high reliability and millimeter-level latency simultaneously if collisions across users frequently happen. In fact, the collision level mainly replies on whether there are sufficient resources for URLLC compared with the network-load, which, however, is typically not known by the base station (BS). To solve this problem, we propose a multi-tier-driven computing framework for URLLC by extracting the network-load information from the access states (success, collision, and idle) of resources across consecutive access slots in a subframe.
%In particular, our multi-tier-driven computing framework consists of three tiers, i.e., network-load learning, network-load prediction and adaptive resource allocation.
%Simulation results corroborate that by exploiting the correlation across multiple access slots, our proposal can achieve more accurate estimation than the baseline scheme. Correspondingly, based on the predictions, our adaptive resource allocation schedule offers a way to enhance the delay QoS requirements coming from different URLLC services simultaneously.
Grant-Free (GF) access has been recognized as a promising candidate for Ultra-Reliable and Low-Latency Communications (URLLC). However, even with GF access, URLLC still may not effectively gain high reliability and millimeter-level latency, simultaneously. %This is because the network load is typically not known to the base station (BS), and thus the collisions become severer when access resource is insufficient. 
This is because the network load is typically time-varying and not known to the base station (BS), and thus, the resource allocated for GF access cannot well adapt to variations of the network load, resulting in low resource utilization efficiency under light network load and leading to severe collisions under heavy network load.
To tackle this problem, we propose a multi-tier-driven computing framework and the associated algorithms for URLLC to support users with different QoS requirements. Especially, we concentrate on $ K $- repetition GF access in light of its simplicity and well-balanced performance for practical systems. In particular, our framework consists of three tiers of computation, namely \emph{network-load learning}, \emph{network-load prediction}, and \emph{adaptive resource allocation}. In the first tier, the BS can learn the network-load information from the states (success, collision, and idle) of random-access resources in terms of resource blocks (RB) and time slots. In the second tier, the network-load variation is effectively predicted based on estimation results from the first tier. Finally, in the third tier, by deriving and weighing the failure probabilities of different groups of users, their QoS requirements, and the predicted network loads, the BS is able to dynamically allocate sufficient resources accommodating the varying network loads. Simulation results show that our proposed approach can estimate the network load more accurately compared with the baseline schemes. Moreover, with the assistance of network-load prediction, our adaptive resource allocation offers an effective way to enhance the QoS for different URLLC services, simultaneously.
\end{abstract}

\begin{IEEEkeywords}
URLLC, grant-free access, multi-tier-driven computing, network-load estimation, adaptive resource allocation 
\end{IEEEkeywords}

\section{Introduction}
Ultra-Reliable Low-Latency Communications (URLLC), along with enhanced Mobile Broadband (eMBB) and massive Machine-Type Communications (mMTC) are the three main application scenarios of the fifth generation (5G) of mobile communications networks. As the 5G had inspired widespread research efforts, some researchers begin to envisage the next steps in wireless networking~\cite{9136591,9360855,9573421}. In beyond 5G (B5G) and 6th generation (6G), URLLC will keep evolving to its advanced version, while still encountering many challenges. The typical quality-of-service (QoS) of URLLC services requires the user-plane latency between the base station (e.g., eNB in 4G and gNB in 5G) and the user to be confined within 1~ms. In the meantime, for transmission of short packets, the reliability needs to be guaranteed with a probability equal to 99.999\%~\cite{3gpp.38.913}. 

\subsection{State-of-the-Art}
The overall latency mainly comes from several factors, including handshake procedures in random access, retransmission in case of collision, scheduling latency introduced by the base station (BS), transmission delay, hardware processing delay at the receiver, etc. Aside from transmission delay and hardware processing delay, which can already be confined within 0.5 through 1 ms currently, even delay caused by the standard handshake procedures for random access~\cite{3gpp.36.213} will inevitably exceed 1~ms. In order to shorten the overall latency, Cheng \textit{et al.}~\cite{9635675} proposed an adaptive block-length transmission framework considering the tradeoff between the transmission delay and the queuing delay. In~\cite{8723547}, Qiao \textit{et al.} derived the maximum throughput that can be supported under statistical queuing delay constraints. In~\cite{8412507}, Gu \textit{et al.} analyzed the effective capacity for machine-type communications with statistical delay constraints. The above research efforts mainly concentrated on the queuing delay. However, the most challenging concerned in URLLC lies in the delay introduced during the random access phase rather than the transmission phase. Moreover, URLLC typically serves the short-packet yet sparse transmissions for each user, the queuing delay of each user’s transmission does not play an essential role in contributing to the overall latency.

To assure millisecond-level latency, URLLC typically employs grant-free (GF) access mode~\cite{9521577,9689965}, where the delay can be significantly shortened by avoiding too many handshake procedures. Typical GF access approaches include the Reactive scheme~\cite{9335255,8877253}, the $K$-repetition scheme \cite{8835946,8877253,9174916}, and the Proactive scheme~\cite{9174916,8877253}, which use redundant transmissions (retransmission and/or repetition transmission) to combat collisions and improve the reliability. In the reactive scheme, retransmission begins if a negative feedback from the BS is received by the user. In the $K$-repetition scheme, each packet is directly and repeatedly transmitted $K$ times within a subframe over $K$ different resource units (e.g., $K$ resource blocks), which can well tradeoff between the retransmission delay and reliability. In the proactive scheme, the BS immediately notifies the user at once upon the successful packet arrival, such that the repetition transmission can stop as early as possible and the occupied random-access resources can be released. 

Based on the existing research and results, it is evident that GF access integrates random access and data transmission together and thus avoiding time-consuming handshakes and shortening waiting delay for BS’s scheduling information. Yet GF access still faces the essential issues in random access, i.e., collisions across users, especially given the fact that the network load of URLLLC is time varying. As the network load is typically not known to the system, the resource allocated for random access of URLLC is fixed. If too many resources are reserved, the utilization efficiency would be very poor; on the contrary, if random-access resources are insufficient, frequent collisions under bursty requests will harm the reliability and prolong the delay by retransmission. Consequently, the key to effectively lower delay and assure reliability is to allocate sufficient resources to URLLC well matching the network load, i.e., the number of active users. Consequently, it is highly desirable to develop network-load estimation techniques and then enable adaptive resource allocation for URLLC, benefiting both latency and reliability quality-of-service (QoS).

%\begin{figure*}  
%	\centering  
%	\setlength{\abovecaptionskip}{-0.15cm}
%	\includegraphics[width=182mm]{System3.pdf}\\  
%	\caption{System model for $ K $-repetition access, load estimation and adaptive resource allocation in URLLC.} 
%	\label{fig:env}  
%	\vspace{-0.5cm}
%\end{figure*}  

There have been some research focusing on network-load estimation in random access, but mainly for machine-to-machine (M2M) communications or mMTC~\cite{2017Traffic,8565923,8880524}. Particularly, \cite{2017Traffic} proposed the traffic-load estimation framework and scheme based on the Markov Chain model. Reference~\cite{8565923} derived the joint PDF of the number of successful and collided transmission attempts. Reference~\cite{8880524} proposed the estimation approach by minimizing the Euclidian distance between the observed preamble access states and the theoretical means. However, it is worth noting that in M2M or mMTC networks, random backoff strategies are applied as response to collisions. As a result, the correlation between access status across adjacent time slots is weakened. In contrast, for URLLC, repetition and immediate retransmission in fact introduce much stronger correlation across adjacent time slots, making estimation approaches for mMTC less inaccurate for URLLC. 

Multi-tier computing has been regarded as an open topic and powerful tool to attain excellent network performance. Multi-tier computing concept can be either fit for task optimization (e.g., scheduling, caching, power allocation, and/or offloading)~\cite{9676649,8713794} or used for network topology partition~\cite{7997056}. We in this paper show that multi-tier computing can be well designed to serve as an effective approach for resolving the aforementioned problems in random access of URLLC. Towards this end, we propose a multi-tier-driven computing framework and the associated algorithms for GF random access in URLLC. The multi-tier-driven computing framework consists of three tiers, namely, \emph{network-load learning}, \emph{network-load prediction}, and \emph{adaptive resource allocation}. 
In this paper, we concentrate on $K$-repetition GF access in light of its simplicity and well-balanced reliability and delay performances, which are highly desirable to practical systems. The idea behind the three-tier computing framework comes from the following principles. The random access resources (e.g., resource blocks) will be randomly selected by users. Consequently, the access states (success, collision, or idle) of resource blocks somehow will carry the information of network load (the active number of access users) in an implicit and hidden manner, which, however, is often neglected. Following this thought, the first-tier computing is designed to learn network-load information from the access states of resource blocks. Then, with the extracted network-load information and the recorded history data, the network load in the coming time slot can be forecast via the second-tier computing. The predicated results will then be injected into the third-tier computing to yield the amount of resources required to accommodate the coming network load, such that QoS assurance for URLLC can be well fulfilled. Also conducted is a set of simulation results to verify the superiority of our proposal compared with the existing baseline approaches. 

\subsection{Contribution}
The main contributions of this paper are summarized as follows: 
\begin{itemize}
	\item We propose the three-tier computing framework for URLLC adopting $K$-repetition access, serving for network-load learning, prediction, and resource allocation, respectively, towards effectively accommodating the varying access load and fulfilling all users’ access with stringent yet differentiated QoS requirements. This framework divides the challenging task, i.e., solving for the resource amount needed to support users’ QoS under the unknown network load, into three tractable tiers
	\item We propose a spectrum of estimation schemes for network-load learning over URLLC based on the access states (success, collision, or idle) of resources blocks, which are suitable for two variant modes of $K$-repetition GF access, termed adjacent-occupation $K$-repetition access and arbitrary-occupation $K$-repetition access. Simulation results demonstrated the superiority of our proposed approach compared with the existing baseline schemes. 
	\item We design an adaptive resource allocation scheme driven by differentiated QoS requirements. In particular, we successfully derive the analytical expressions of access -failure probability within 1~ms for the two $K$-repetition access schemes as a function of network load. Then, the system can precisely allocate resources to different types of services according to their QoS requirements and predicated traffic loads. We validate our proposed algorithms via abundant simulations under various network conditions. 
\end{itemize}

\subsection{Structure}
The rest of the paper is organized as follows. Section II describes the system model and proposes the three-tier computing framework for URLLC. Section III proposes a spectrum of network-load estimation schemes in the first-tier computing for $K$-repetition access based URLLC.  Section IV concentrates on the second-tier computing and presents the network-load prediction schemes. Section V derives the third-tier computing to develop the adaptive resource allocation scheme. Section VI presents the simulation results. The paper concludes with Section VII.

\section{System model}

\subsection{System Description}

We consider a multi-user network for URLLC which is composed of $ N_{\mathrm {al}} $ users and one base station (BS).
These users can be in only two states, i.e., \textit{active} and \textit{inactive} states. The number of active users is denoted by $ N_{\mathrm {tr}} $.
The users access and transmit in a synchronized but Grant-Free (GF) manner, which is coordinated by the BS.
It is well-known that URLLC requires high successful access probability and low latency, which heavily relies on whether  there are sufficient resources for GF access compared to $ N_{\mathrm {tr}} $. However, $ N_{\mathrm {tr}} $ is typically unknown to the BS, thus imposing the major hurdle to release GF access potentials.
In this paper, we attempt to estimate $ N_{\mathrm {tr}} $, then enabling adaptive resource allocation to better support URLLC with assured QoS.

\subsection{The Multi-Tier-Driven Computing Framework}
\begin{figure}  
	\centering  
	\includegraphics[width=9cm]{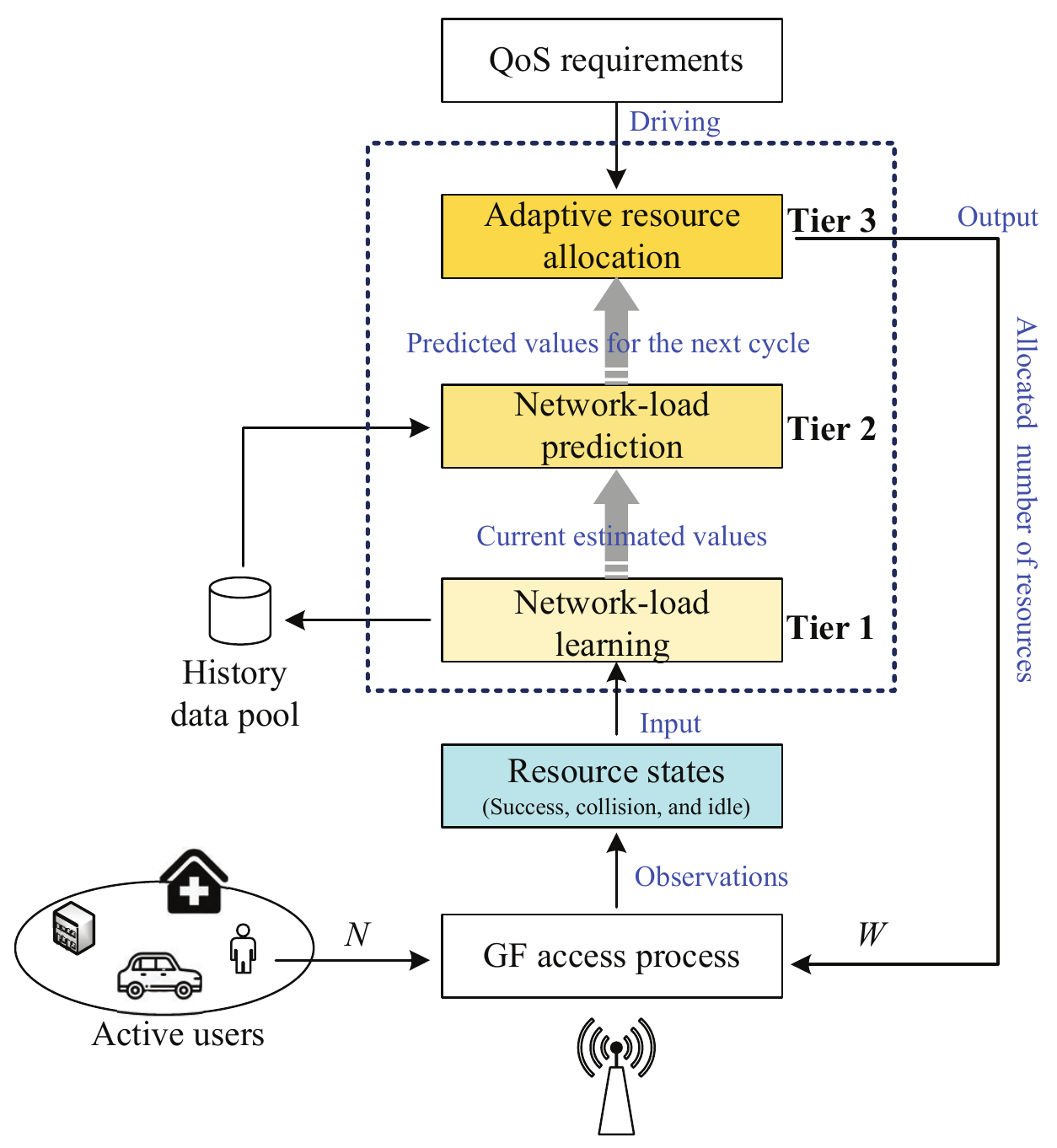}\\  
	\caption{The multi-tier-driven computing framework. }
	\label{multi-tier}  
\end{figure}

The access states (success, collision, or idle) of resource blocks can include hidden information of network-loads, which, however, are typically neglected.
In this paper, we propose a multi-tier-driven computing framework consists of three tiers, i.e., network-load learning, network-load prediction, and adaptive resource allocation, in order to assure the QoS requirements of URLLC applications.
The network-load learning is implemented via a Markov-chain model based estimation technique, and thus in this paper we use the terminologies \emph{network-load learning} and \emph{network-load estimation} interchangeably.
As shown in Fig.~\ref{multi-tier}, firstly the BS can get the knowledge of resource states in every slot (success, collision, and idle). Based on these observations, we can get the estimated number of current active users via the first tier.
The estimated values will be recorded in the history data pool and also utilized as the input together with a selected series of history data to the second tier.
Based on the prediction values, the third tier will adaptively allocate resources driven by different QoS requirements.
In particular, we propose several network-load estimation schemes for adjacent-occupation $ K $-repetition and arbitrary-occupation $ K $-repetition, respectively.
We also employ ARIMA model to achieve accurate and timely predictions. 
With the assistance of analytical formulations towards access failure probability, 
we design a resource negotiation algorithm driven by QoS requirements.

\subsection{K-Repetition Access and Transmissions}

\begin{figure}  
	\centering  
	\includegraphics[width=9cm]{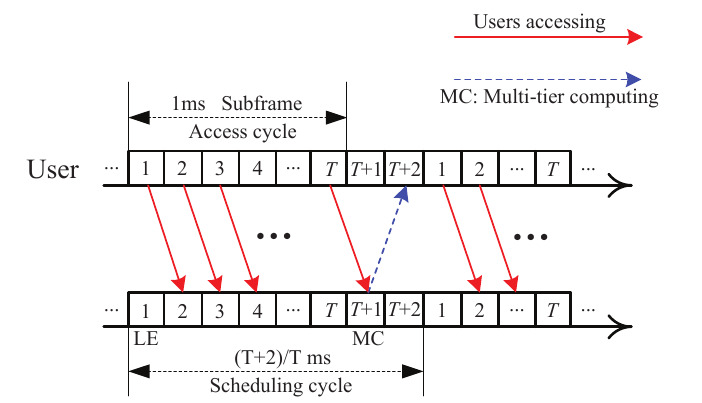}\\  
	\caption{The illustration of $ K $-repetition, access cycle, and scheduling cycle.} 
	\label{access_cycle}  
\end{figure}

We will use resource block (RB) to denote a generalized concept in the following description, e.g., a RB includes 12 consecutive subcarriers in LTE and 5G NR system. 
The BS divides time into consecutive subframes. \footnote{In practical networks, a number of consecutive subframes typically together form a frame.} Each subframe is dedicated for a GF \textit{access cycle} and is further divided into $ T $ slots \footnote{It is often referred to as mini-slot in 5G. But in this paper, we use the term slot in short for simplicity.} with equal length. Between two adjacent subframes, two slots are inserted for the BS which are used to broadcast the available resource information for next cycle to users. For presentation convenience, we call an access cycle and the two followed slots as a \textit{scheduling cycle}, as shown in Fig. ~\ref{access_cycle}.
Several GF access approaches have been proposed by researchers, and typical ones include Reactive, $ K $-Repetition, and Proactive \cite{8877253}. In this paper, we mainly concentrate on the $ K $-repetition scheme, in light of its simplicity and well-balanced performances.
In the $ K $-repetition scheme, the BS permits every user to repeatedly access and transmit $ K $ times in consecutive slots (i.e., \textit{adjacent-occupation}) or arbitrary but different slots (i.e., \textit{Arbitrary-occupation}) of an access cycle.
In every access slot, each user is permitted to occupy only one RB.
If a RB is occupied by only one user in a slot, this access is successful.
Otherwise, if there are two or more users occupying the same RB in a slot, they all happen to collide and fail.
Only when all the replicas fail in a cycle, this user needs to retry in the following cycles.

\subsection{Statistical Feature of URLLC}
\label{traffic_patterns}
The general URLLC requirement for one transmission of a packet is 99.999\% reliability and latency within 1ms.
A typical type of URLLC application is in the scenario of sophisticated industrial production and controlling, in which the machine-type devices, e.g., sensors and controllers, always call for communication services with high reliability and low latency.
The industrial internet of things (IIoT) applications with URLLC requirements can be classified into two main types of use cases~\cite{3gpp.122.261,8663990}: motion control and discrete automation.
For motion control with continuous and stable data transmission, e.g., automatic production machine tools and 3D printers, this type of services can be regarded as uniform and periodic pattern~\cite{8638959}, which has stable access intensity.
While for discrete automation, it always has difficulty in accurately predicting when the packets flow will arrive, especially in vehicle to everything (V2X) communications~\cite{8269135}. 

Fortunately, the internal activation order of a batch of bursty devices can follow some distributions.
In particular, the Beta distribution is one of the suitable types to model this process~\cite{8269135, 8761430,weerasinghe2020priority}, in which all of the devices access to the BS in a deterministic order during a limited period~\cite{3gpp.37.868}.
The Beta distribution in bursty traffic has been defined in~\cite{3gpp.37.868}:
\begin{equation}
	N_i=N\int_{t_{i-1}}^{t_i}p(t)dt,\label{Beta}
\end{equation}
in which $ N_i $ denotes the number of users in the $ i $th access cycle of the whole event duration. 

In this model, $ t_i-t_{i-1}$ is equal to the length of a complete scheduling cycle.
Supposing that each subframe is divided into 8 slots in this paper, if the duration of Beta distribution is set as 12.5~ms, we have $i\in\left\{\mathbb{Z}\mid\left[1,10\right]\right\} $ considering two scheduling slots in every scheduling cycle.

The $ p(t) $  in (\ref{Beta}) is derived by:
\begin{equation}
	p(t)=\frac{{{t}^{\alpha -1}}{{(T-t)}^{\beta -1}}}{{{T}^{\alpha +\beta -1}}\mathcal{B}(\alpha ,\beta )}
\end{equation}
in which $\mathcal{B}(\alpha ,\beta )=t^{\alpha-1}(1-t)^{\beta-1}$ with $ \alpha=3 $ and $ \beta=4 $ typically.
We will use $ \mathcal{B}(N,T) $ to represent that there are $ N $ users in total to access in $ T $~ms with bursty traffic pattern.
%In this paper, we will discuss these two traffic patterns and also their fusion type, and refer to them as uniform traffic, bursty traffic and hybrid traffic respectively.
The parameters used in this paper are summarized in Table \ref{table}.
\begin{table}[!t]
	\centering
	\caption{Parameters and Notations}
	\label{table}
	\rowcolors{2}{gray!25}{white}
	\begin{tabular}{m{1.2cm}<{\centering}| m{6cm}}
		\hline  % 顶部线
		%\rowcolors {gray!50}	
		Parameters & Descriptions \\ 
		\hline$ N_{\mathrm{al}} $&The total number of users\\
		\hline
		$  N_{\mathrm{tr}} $&The number of active users who attempt to access\\
		\hline$ \widehat{{N}}_{\mathrm{tr}} $&The estimation of $  N_{\mathrm{tr}} $\\
		\hline$ N $&The test number of users for hypothesis in ML\\
		\hline$ E $&The prediction of the number of users in next 1ms\\
		\hline$  \widehat {n}_t $&The estimation of the number of users who access in the $ t $-th slot\\
		\hline	$ {\mathbf n} $&The load estimation vector:  $ {\mathbf n} =(  \widehat {n}_1 , \widehat{n}_2 , \ldots,\widehat{n}_T )^AT$\\
		\hline	$ \phi_r  $&The number of users beginning their first access in the $r_{th} $ slot\\
		\hline	$ {\boldsymbol \phi} $&The starting vector:  $ {\boldsymbol \phi} = ({\phi_1} , {\phi_2} , \ldots, {\phi_{T-K+1}} )^ T  $\\
		\hline $A$&The number of RBs each of which is selected by only one user to access, suggesting successful access\\
		\hline $ B $& The number of RBs each of which is selected by two or more users to access, suggesting collision\\
		\hline $ C $&The number of RBs each of which is selected by any users, suggesting idle states\\
		\hline	$(A,B,C)$&The Markov state denoting the number of success RBs, collision RBs and empty RBs, respectively \\
		\hline	${\mathbf P}$&Markovian transition matrix\\
		\hline	${\mathbf S}$&Markovian space for all RBs' states\\
		\textit{W}&The total number of available RBs\\
		\hline	$ { K} $&The parameter for $ K $-repetition: the number of consecutive slots a user attempt to send access signals\\
		\hline	$ { T} $&The total number of slots within a subframe\\		
		\hline % 底部线		
		
	\end{tabular}
	%\vspace{-0.5cm}
\end{table}

\section{Network-Load Estimation }
The adjacent-occupation $ K $-repetition has strict requirements on the slot selection for single user, while the arbitrary-occupation $ K $-repetition allows a user to access in optional but different $ K $ slots and doesn't emphasize the adjacency.
We propose two estimation schemes for adjacent-occupation $ K $-repetition in Section~\ref{section-NLE-A} and \ref{MS-ML}, and one estimation scheme for arbitrary-occupation $ K $-repetition in Section~\ref{MS-MLD}, respectively. We list the suitability of each scheme in Table~\ref{matching}.
\begin{table}[t]
	\vspace{-0.3cm}   
	\centering
	\caption{The suitability of network-load estimation schemes}   
%\begin{tabular}{m{0.85cm}<{\centering}m{2.45cm}<{\centering}m{1.7cm}<{\centering}m{0.7cm}<{\centering}m{1cm}}    
		\begin{tabular}{cccc}    
		\toprule    
\multirow{2}{*}{$ K $-repetition type}  & \multicolumn{3}{c}{Estimation schemes} \\
\cline{2-4}
& SS-ML-LS & MS-MLI & MS-MLD\\
\cline{1-4}
Adjacent-occupation& \checkmark & \checkmark &  \\
\cline{1-4}
Arbitrary-occupation& & & \checkmark \\
		\bottomrule   
	\end{tabular}\label{matching}
\end{table}

%Introduce the four kinds of estimation schemes briefly which should be fit for different Krep schemes. 

%Based on the above problem, we believe that there is an urgent need for BS to figure out the number of active UEs after  every 1ms on the basis of the known data, without which it cannot reasonably allocate the RBs for next 1ms. Hence, this section proposes a network load estimation framework including four novel schemes aimed at different types of $\rm K_{rep} $. More specifically, the scheme \textit{C.} below is fit for the Optional $\rm K_{rep} $, the scheme \textit{D.} is fit for  the Adjacent $\rm K_{rep} $, while the scheme \textit{B.} is fir for both. 

%And the work [] has already proposed a estimation scheme based on the minimum likelihood of variance, however, it applies to the single access of massive UEs. To compare this scheme with ours, it is  transformed as  the scheme \textit{A.} below, which is also fit for both of the $\rm K_{rep} $.

\subsection{Single-Slot Maximum Likelihood with Least Squares Estimation}
\label{section-NLE-A}
In the Single-slot Maximum Likelihood with Least Squares Estimation (SS-ML-LS),  we concentrate on the relationship between the number of users beginning their first access and total active users, which are denoted as $ \phi_r (1\leqslant r\leqslant { T-K+1})$ and $ n_t (1\leqslant t\leqslant T)$, respectively. The $ \phi_r $ denotes the number of users who begin  their first access in the $ r $th slot, and the $ n_t $ denotes the total number of active users in the $ t $-th slot. Thus we use this equation set below to describe the relationship:
\begin{equation}
	n_t = \left\{
	\begin{array}{cl}
		\displaystyle\sum_{k=1}^{t}\phi_{k}, &\mbox{if} ~ 1\leq t < K;\\
		\displaystyle\sum_{k=t-K+1}^{t}\phi_{k}, &  \mbox{if} ~K\leq t \leq  T-K+1;\\
		%\phi_{t-{(K-1)}}+\phi_{t-{ (K-2)}}+\cdots+\phi_{t}
		\displaystyle\sum_{k=t-K+1}^{T-K+1}\phi_{k}, &  \mbox{if} ~T-K+1< t \leq T,\\
	\end{array}
	\right.
	\label{fangcheng}
\end{equation}
note that users should initiate the first access before and excluding $ T-K+2 $ th slot, otherwise they are not able to finish $ K $ times in an access cycle. 

To obtain the total amounts of users in a cycle, firstly we aim at the number $ n_t $ of users in every slot.
We use $ A $, $ B $ and $ C $ to denote the number of success RBs, collision RBs, and empty RBs respectively observed by the BS in every slot.
Definitely, here is  ${ A+B+C} = W$. Though this is a established fact made by users' RB choices, it is a new perspective to view the process as a Markov model with $ N $-steps transition, in which every user's RB choice is corresponding to one step of Markov transition. 
%Maximum A Posteriori Probability Estimate (MAP) can be used to modeling as follows:
%	\begin{equation}
%	\begin{split}
%		\widehat{N}_{\mathrm{tr}}&= \arg \max \limits_{N}P \big(N \,| \, ({ A,B,C}) \big)\\
%		&= \arg \max\limits_{N} \dfrac{P \big(({ A,B,C}) \, | \, N \big)P(N)}{P({ A,B,C})}\\
%		&=\arg \max\limits_{N} {P \big(({ A,B,C}) \, | \, N \big)P(N)}\,,
%	\end{split}
%\end{equation}

%\noindent where  $ \widehat{N}_{\mathrm{tr}} $ denotes the estimation result, and Baye's law converts the posterior probability $ P \big(N \,| \, ({ A,B,C}) \big) $ into the product of likelihood function $ P \big(({ A,B,C}) \, | \, N \big) $ and the prior distribution of parameter $ P(N) $. Note that $ P({ A,B,C}) $  doesn't depend on \textit{N}.

Thus we employ Maximum Likelihood Estimate (ML)  to find the most likely $ N $ for Markov model, which is denoted by $ \widehat{N}_{\mathrm{tr}} $:
%However, it is difficult for BS to obtain $ P(N) $ in advance, which is related to the user distribution. Thus we turn to employ another estimation method, namely Maximum Likelihood Estimate (ML), with the Markov model:
	\begin{equation}
	\begin{split}
		\widehat{N}_{\mathrm{tr}} &= \arg \max \limits_{N} P \big(({ A,B,C}) \, | \, N \big)\\
		&= \arg \max\limits_{N} {\mathbf P}_{({ 0,0,W}) \, \rightarrow \, ({ A,B,C})}^{N}\,,\label{zong}
	\end{split}
\end{equation}

\noindent where ${\mathbf P}$ represents the transition matrix of Markov model, superscript $ N $ denotes the transition steps for hypothesis, and the subscript  $ ({ 0,0,W}) \, \rightarrow \, ({ A,B,C}) $ shows the initial state and the terminal state in Markov model, respectively. 

The state space ${\mathbf S}$ of Markov model containing all the possible states can be formulated with following described rules:
	\begin{equation}
	\left\{
	\begin{array}{lr}
		a+b+c = W;\vspace{3pt}\\
			0 \leqslant a \leqslant N;\vspace{3pt}\\
			 0 \leqslant b \leqslant \frac{N}{2};\vspace{3pt}\\
			  0\leqslant c \leqslant W,
	\end{array}
	\right.
\end{equation}
\noindent where $ (a,b,c) $ represent the possible states, and other $ a $, $ b $ and $ c $ obeying the rules are not allowed. For state $(A,B,C)$ in ${\mathbf S}$, we define the transition probabilities from $(A,B,C)$  to other states  as:
	\begin{equation}
	\left\{
	\begin{array}{lcc}
		P_{(A,B,C) \rightarrow (A,B,C)} &=& \frac{B}{W};\vspace{3pt}\\
		P_{(A,B,C) \rightarrow ({A+1},B,{C-1})} &=& \frac{C}{W};\vspace{3pt}\\
		P_{(A,B,C) \rightarrow ({A-1},{B+1},C)} &=& \frac{A}{W};\vspace{3pt}\\
		P_{(A,B,C) \rightarrow \mathrm{others}} &=& 0.				 	
	\end{array}\label{zhuanyi}
	\right.
\end{equation}

The probabilities that state $(A,B,C)$  transfers into above three states can be explained by:
%Except  above three states, the probability of transitioning to another state is zero. And these states that the state $ (A,B,C) $ may change into by once more transition can be respectively derived from:
\begin{itemize}	
	\item   The new user randomly chooses a RB belonging to the collision RBs with probability $ \frac{B}{W} $, and both the number of success and collision RBs don't change;
	\item	The new user randomly chooses a RB belonging to the empty RBs with probability $ \frac{C}{W} $, thus the number of success RBs increases;
	\item   The new user randomly chooses a RB belonging to the success RBs exactly with probability $ \frac{A}{W} $, causing collision between this user and the other user who has chosen this RB, hereafter, the number of success RBs decreases while the number of collision RBs increases.
\end{itemize}
	 
Substituting (\ref{zhuanyi}) into (\ref{zong}), we can obtain the most likely number of users in the $ t $-th slot which is denoted by $n_{t}$ $(1 \leqslant t \leqslant {\rm T})$.

 We further define $ {\boldsymbol \phi}=({\phi_1}, {\phi_2} , \ldots, {\phi_{T-K+1}})^T$,  $ \mathbf{n}=({n_1}, {n_2} , \ldots, {n_T})^T $, and use $ \boldsymbol{\Omega} $ to denote a $ \left( T\times \left(T-K+1 \right) \right) $ matrix. Then we have an overdetermined equation as follows:
\begin{equation}
		\boldsymbol{\Omega}  {\boldsymbol \phi}  = {\mathbf n},
			\label{LS}
\end{equation}
where $ \boldsymbol{\Omega} $ can be derived by:
\begin{equation}
		\boldsymbol{\Omega}=\left[
		\begin{array}{c}
			  \boldsymbol{\Gamma}_1 \\
			 \boldsymbol{\Gamma}_2\\
			 \boldsymbol{\Gamma}_3 \\
		\end{array}\right].\\
\end{equation}

We can obtain $ \boldsymbol{\Gamma}_1 $, $ \boldsymbol{\Gamma}_2 $, and $ \boldsymbol{\Gamma}_3 $ by:
\begin{equation}
	\begin{aligned}
		\boldsymbol {\Gamma}_1=&\left[
		\mathbf{1}^{\mathrm{low}}_{{ \left(K-1\right)\times\left(K-1\right)}}~~ \mathbf{0}_{{ \left(K-1\right)\times\left(T-2(K-1)\right)}}
		\right];\\
		\boldsymbol{\Gamma}_3=&\left[
		\mathbf{0}_{{ \left(K-1\right)\times\left(T-2(K-1)\right)}} ~~ \mathbf{1}^{\mathrm{up}}_{{ \left(K-1\right)\times\left(K-1\right)}}
		\right],\\		
	\end{aligned}  
\end{equation}
where $ \mathbf{1}^{\mathrm{low}}_{{ \left(K-1\right)\times\left(K-1\right)}} $ is a $ (K-1)\times(K-1) $ lower triangular matrix in which the main diagonal and all entries below the main diagonal are equal to 1, $ \mathbf{1}^{\mathrm{up}}_{{ \left(K-1\right)\times\left(K-1\right)}}$  denotes a $ (K-1)\times(K-1) $ upper triangular matrix in which the main diagonal and all entries above it are equal to 1, and $ \mathbf{0}_{{ \left(K-1\right)\times\left(T-2(K-1)\right)}} $ represents a $ (K-1)\times(T-2(K-1)) $ matrix in which all entries are equal to 0.  $ \boldsymbol{\Gamma}_2 $ represents a $ (T-2(K-1))\times(T-(K-1)) $ which can be derived by:
\begin{eqnarray}
	\boldsymbol{\Gamma}_2 \!=\!\left[\!\!\!\!\!\!
	\begin{array}{c}
		\begin{tabular}{p{0.6cm}<{\centering}p{0.6cm}<{\centering}p{0.6cm}<{\centering}p{0.6cm}<{\centering}p{0.6cm}<{\centering}p{0.6cm}<{\centering}p{0.6cm}<{\centering}}
			1 & $\cdots$ & 1 & 0 &  $\cdots$ & $\cdots$ & 0\\
		\end{tabular}	\vspace{-10pt}\\
		\hspace{-120pt}\underbrace{~\quad\quad\quad\quad\quad\quad~}_{K~ \mbox{\footnotesize times}}\vspace{3pt}
		\\
		\begin{tabular}{p{0.6cm}<{\centering}p{0.6cm}<{\centering}p{0.6cm}<{\centering}p{0.6cm}<{\centering}p{0.6cm}<{\centering}p{0.6cm}<{\centering}p{0.6cm}<{\centering}}
			0 & 1 & $\cdots$ & 1 & 0 & $\cdots$ & 0\\
		\end{tabular}\vspace{-10pt}\\
		\hspace{-60pt}\underbrace{~\quad\quad\quad\quad\quad\quad~}_{K~ \mbox{\footnotesize times}}\vspace{0pt}
		\\
		\begin{tabular}{p{0.6cm}<{\centering}p{0.6cm}<{\centering}p{0.6cm}<{\centering}p{0.6cm}<{\centering}p{0.6cm}<{\centering}p{0.6cm}<{\centering}p{0.6cm}<{\centering}}
			$\vdots$ & $\vdots$ &$\vdots$ &$\ddots$ &$\vdots$ &$\vdots$ & $\vdots$\vspace{5pt} \\
		\end{tabular}\\
		\begin{tabular}{p{0.6cm}<{\centering}p{0.6cm}<{\centering}p{0.6cm}<{\centering}p{0.6cm}<{\centering}p{0.6cm}<{\centering}p{0.6cm}<{\centering}p{0.6cm}<{\centering}}
			0 & $\cdots$ & $\cdots$ & 0 & 1 &  $\cdots$ & 1
		\end{tabular}\vspace{-10pt}\\
		\hspace{120pt}\underbrace{~\quad\quad\quad\quad\quad\quad~}_{K~ \mbox{\footnotesize times}}
	\end{array}
	\!\!\!\!\!\!\right].
\end{eqnarray}
%_{(K-1) \times (K-1)}
%_{(K-1) \times (T-K+1)}
%	

After obtaining the estimation $ \mathbf n $ with ML, we can search an approximate solution for (\ref{LS}):
	\begin{equation}
	\begin{array}{c}
	{\boldsymbol \phi} \quad = \quad\left( {\boldsymbol{\Omega}^T}\boldsymbol{\Omega} \right)^{-1} {\boldsymbol{\Omega}^ T} {\mathbf n}. 
	\end{array}
\end{equation}
Finally, we obtain the total number of users with $ \phi_r $ as follows:
	\begin{equation}
	\widehat{{N}} _{\mathrm{tr}}= \sum_{r=1}^{T-K+1} \phi_r\,.
\end{equation}

\subsection{Multi-Slot Maximum Likelihood Indirect Estimation for Adjacent-Occupation $ K $-Repetition }
\label{MS-ML}
In the Multi-Slot Maximum Likelihood Indirect Estimation (MS-MLI), we consider solving this problem in multi-slot  which turns out to be more accurate.  Firstly,  we denote the access states of RBs in the \textit{t}-th slot as $ (A_t, B_t, C_t) $,  and then we use three vectors to record these states respectively, i.e.,  $ \mathbf A$, $ \mathbf B$ and $ \mathbf C$. For example, $ \mathbf A= ({A_1}, {A_2} , \ldots,  A_T )^T$. For every test number $ N $ of users, it determines the space of all possible vector $ {\boldsymbol \phi}$ accordingly, denoting the amount of users beginning their first access in every slot. The probability mass function (PMF)of every $ {\boldsymbol \phi}$ can be defined with this equation:
	\begin{equation}
	\begin{split}
		Pr \left \{ \boldsymbol \phi \, | \, N \right \} =\dfrac{1}{{ (T-K+1)}^N} \; \prod_{r=1}^{T-K+1} \binom{N- \sum_{j=0}^{r-1} \phi_j}{\phi_r} \,.
	\end{split}\label{PMF}
\end{equation}

Considering the relationship between a specific $ \boldsymbol \phi $ and $ \mathbf n $ has been derived by (\ref{fangcheng}), next we calculate the transition probability from the initial state $ (0,0,W) $ to the terminal state $ (A_t, B_t, C_t) $ with $ n_t $ steps according to $ \mathbf n $, which is similar to (\ref{zong}). The difference is that here $ n_t $ is definite under the restriction of each  $ {\boldsymbol \phi}$,  not a  test number. Finally,  we have the probability of the test number \textit{N} of users with Total Probability Theorem, and the result of ML can be formulated by:
	\begin{equation}
	\begin{split}
		\widehat{{N}} _{\mathrm{tr}}&= \arg \max \limits_{N} P \bigl ( \left ({\mathbf A}, {\mathbf B},  {\mathbf C} \right) \, | \, N \bigr )\\
		&= \arg \max \limits_{N}  \sum_{\boldsymbol\phi} \;Pr \biggl \{ \bigl({\mathbf A}, {\mathbf B},  {\mathbf C} \bigr )  \, | \, {\boldsymbol n \left({\boldsymbol \phi} \right)} \biggr \} \; Pr \left \{ \boldsymbol \phi \, | \, N \right \}\\
		&= \arg \max \limits_{N} \sum_{\boldsymbol \phi} \; \biggl\{ \prod_{t=1}^{T} Pr \bigl ( \left (A_t,B_t,C_t \right ) \, | \, n_t \bigr ) \biggr\} \; Pr \left \{ \boldsymbol \phi \, | \, N \right \}  \label{zong3} \,.
	\end{split}
\end{equation}

Thus we infer this scheme as an indirect estimation in that it utilizes mediate information, i.e., the number of users who begin their first access in every slot.

\begin{figure*}[htbp]
	\centering
	\subfigure[]{		
		\includegraphics[width=8cm]{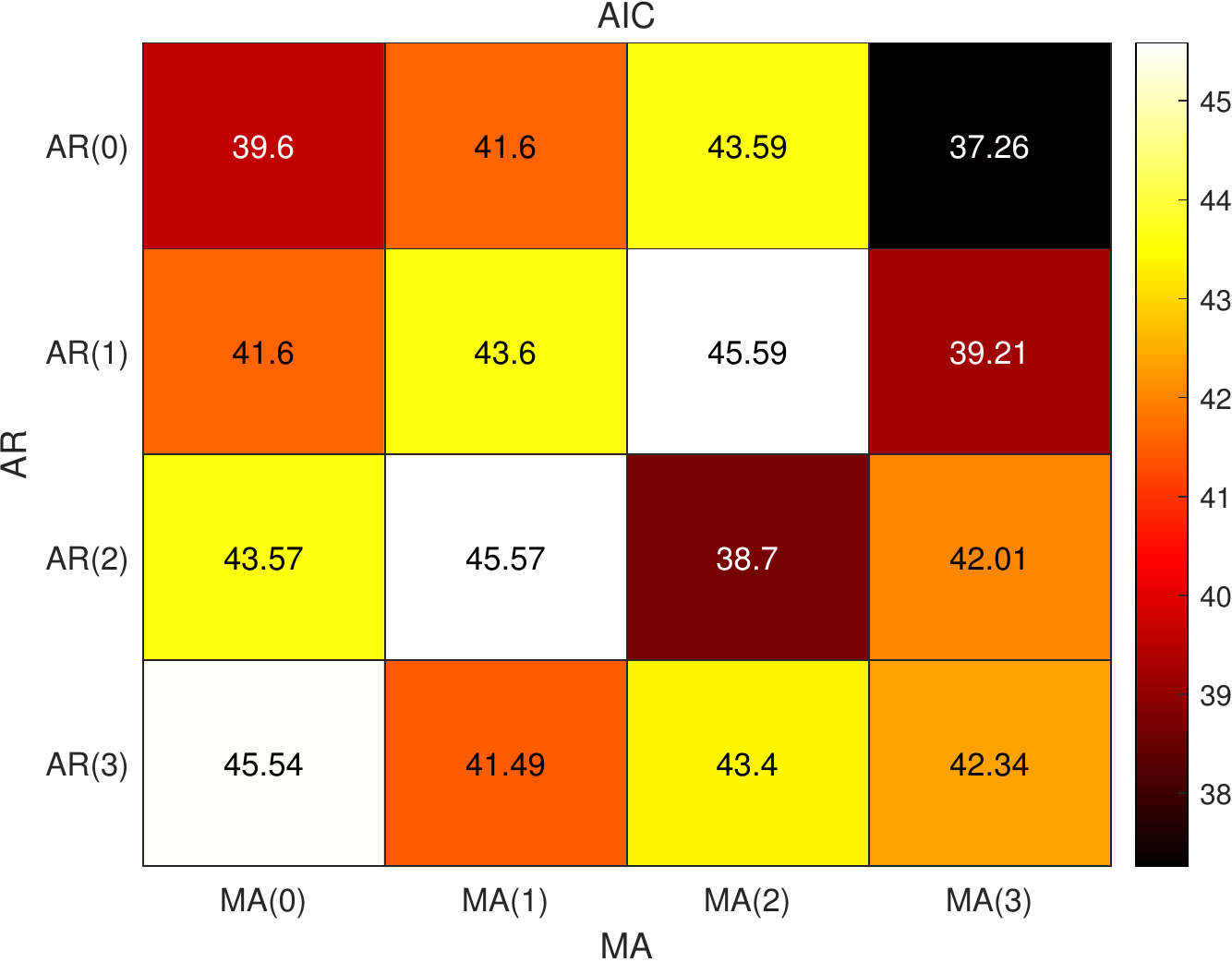}
		\label{AIC}
	}%
	%\hspace{-0.5cm}	
	\subfigure[]{
		\includegraphics[width=8cm]{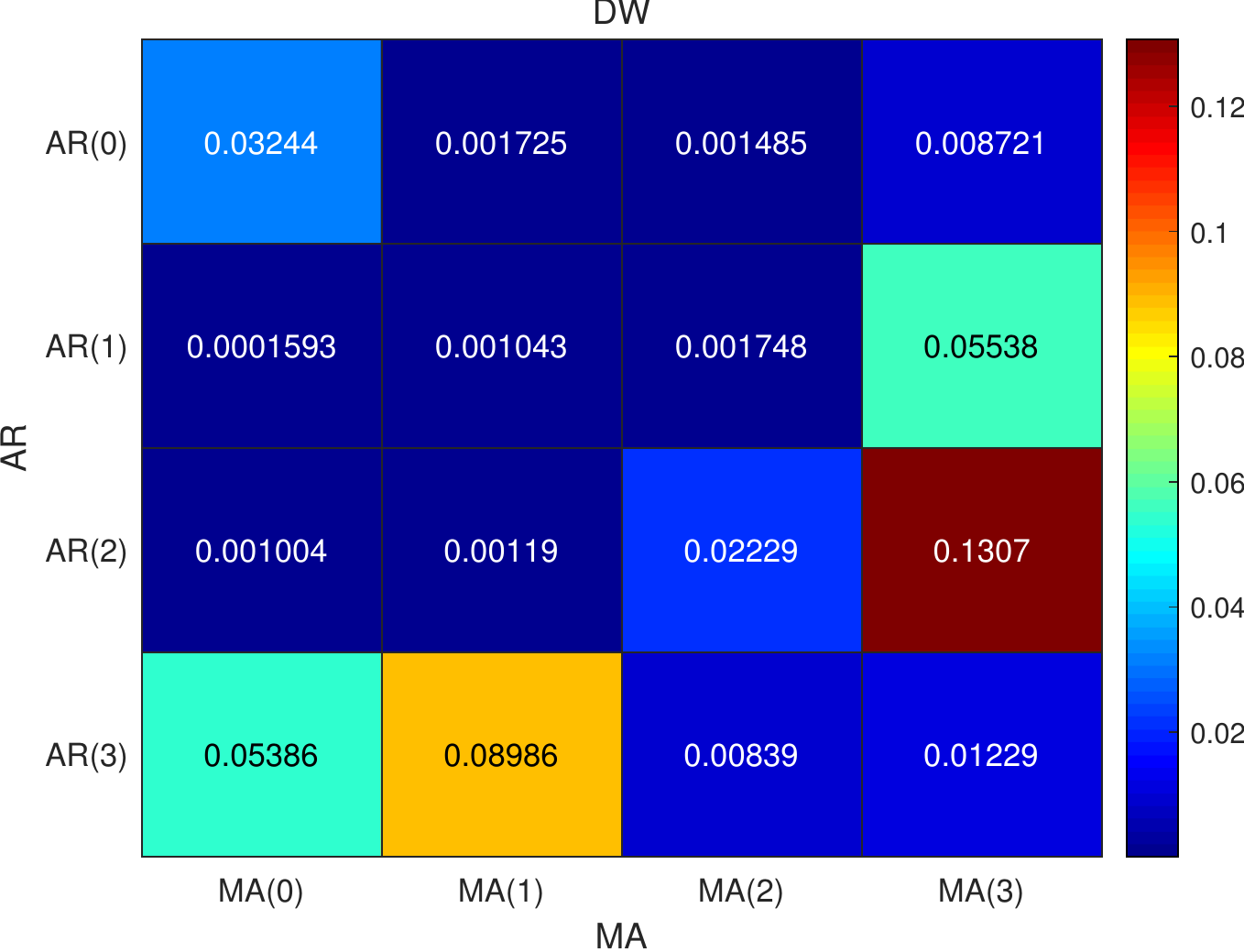} 
		\label{DW}
	}
	\centering
	\caption{The prediction performances of ARIMA model using different metrics. (a) uses the AIC metric. (b) uses the DW metric, and all the values have been minus 2 and taken absolute. }
	%(c) shows the predicted loads with congestion attack. (d) shows the predicted loads with false data dissemination.
	\label{pqchoose}
\end{figure*}

\subsection{Multi-Slot Maximum Likelihood Direct Estimation for Arbitrary-Occupation $ K $-Repetition}
\label{MS-MLD}
In the Multi-Slot Maximum Likelihood Direct Estimation (MS-MLD), a single user is still considered as a transition step in Markov model, but will cause the states of $ K $ RBs to change at one time.
The state space ${\mathbf S}$ of Markov model containing all the possible states can be formulated with following described rules:
\begin{equation}
	\left\{
	\begin{array}{lr}
		a+b+c = WT;\vspace{3pt}\\
		0 \leqslant a \leqslant N;\vspace{3pt}\\
		0 \leqslant b \leqslant \frac{N}{2};\vspace{3pt}\\
		0\leqslant c \leqslant W,
	\end{array}
	\right.
\end{equation}
\noindent where $ (a,b,c) $ represent the possible states, and other $ a $, $ b $ and $ c $ obeying the rules are not allowed in ${\mathbf S}$.

For an arbitrary state $ (A, B, C) $ in ${\mathbf S}$, we use $ ( \widetilde{A}, \widetilde{B}, \widetilde{C}) $ to denote the possible next state, which must satisfy the following relationships:
\begin{equation}
		\left\{ 
			\begin{array}{cl}
			A-\text{min}\left( K,A \right) \leqslant \tilde{A} \leqslant A+\text{min}\left( K,C \right);  \\
			B \leqslant  \tilde{B} \leqslant B+\text{min}\left( K,A \right);  \\
			C-\text{min}\left( K,C \right) \leqslant \tilde{C} \leqslant C .
	\end{array}
\right.\label{daxiao}
\end{equation}
Considering the rule of arbitrary-type repetition, we can also have:
\begin{equation}
	\left\{ \begin{array}{lr}
		x=\widetilde{B}-B;  \\
		y=C - \widetilde{C};\\
		z=K-\left( x+y \right),  \\
	\end{array} \right.
	\label{xyz}
\end{equation}
where $ x,y,z \geqslant 0 $.
It can be thought that the single user chooses $ x $ RBs belonging to $ A $ success RBs and causes the number of collision RBs to increase from $ B $ to $ \widetilde{B} $; chooses $ y $ RBs belonging to $ C $ idle RBs and causes the number of idle RBs to reduce from $ C $ to $ \widetilde{C} $; chooses $ z $ RBs belonging to $ B $ collision RBs respectively.
Thus for $ ( \widetilde{A}, \widetilde{B}, \widetilde{C}) $ calculated by (\ref{daxiao}), the (\ref{xyz}) will filter suitable states again.

The transition probability can be formulated as follows:
\begin{equation}
	\left \{
	\begin{array}{lr}
		P_{(A,B,C) \, \rightarrow \, (\widetilde{A},\widetilde{B},\widetilde{C})} =
			\dfrac{\binom{A}{x}\,\binom{C}{y}\,\binom{B}{z}}{\binom{ WT}{ K}}\\
		P_{(A,B,C) \, \rightarrow \, others} \ \;= 0
	\end{array}
	\right.,\label{14}
\end{equation}
in which $ x\in[0,A]$,  $ y\in[0,C] $, and $ z\in[0,B] $.

Thus we can calculate the probabilities of transferring from the initial state $ (0,0,WT) $ to the final observed state via $ N $-steps, and the estimated number $ \widehat{{N}} _{\mathrm{tr}} $ of active users is related to the maximum one, which can be depicted follows:
	\begin{equation}
	\begin{split}
		\widehat{N}_{\mathrm{tr}} &= \arg \max \limits_{N} P \big(({ A,B,C}) \, | \, N \big)\\
		&= \arg \max\limits_{N} {\mathbf P}_{({ 0,0,WT}) \, \rightarrow \, ({ A,B,C})}^{N}\,.
	\end{split}
\end{equation}

Thus we refer this scheme as a direct estimation because it directly considers each user's $ K $ choices and the corresponding transition of resource states, without benefited by the information of vector $ {\boldsymbol \phi}$ which represents the number of users who begin their first access in every slot.

\section{Network-load Prediction}
Though we have mentioned that it's unrealistic to accurately predict for bursty traffic when the occasional event will happen, the prediction module should have ability to timely response once it detects the beginning, which is the critical basis of resource allocation.
There are some common schemes of time series prediction, including simple equal, moving average, exponential smoothing, and machine learning. 
The simple equal scheme assumes that the next expected value is equal to the current observed value, which always falls behind the real change.
The machine learning schemes typically require training process, thus will inevitably raise challenges in model choosing and time complexity.

Then we employ the auto-regressive integrated moving average (ARIMA) model~\cite{box2015time}, which simultaneously considers the past observations and random errors,
to sense and predict the bursty event enabled by the estimated real-time network-loads.
The training sequence is generated by $ \mathcal{B}(100,20) $.
Firstly, we evaluate whether the series of Beta distribution in (\ref{Beta}) are stationary with Augmented Dickey-Fuller, and after making second-order difference, they proves to be stationary.
An ARIMA $ (p,d,q) $ model can be formulated as follows:
\begin{equation}
	\begin{array}{cl}
		\widetilde {n}_t=c+\sum\limits_{i=1}^{p}\phi_i n_{t-i}
		+\sum\limits_{j=1}^{q}\theta_j \varepsilon_{t-j}+\varepsilon_t,
	\end{array}	
\end{equation}
in which $ \widetilde {n}_t $ denotes the predicted value in the $ t $-th point, while $ n_{t-i} $ and $ \varepsilon_{t-j} $ represent the observed values and random errors in the history respectively. 
The random errors are assumed as following the Gaussian distribution, and we will verify its rationality via DW test below.
The $ p $, $ d $, and $ q $ denote the model parameters of past observed item, difference item and error item, respectively.
Note that input training data should make second-order difference, and we have $ d=2 $.

For determining the values of $ p $ and $ q $, we generate a group of data following the Beta distribution,
and employ the Akaike information criterion (AIC) and Durbin-Watson (DW) test to evaluate the model's performances.
DW test can measure the first-order autocorrelation of residuals in regression analysis, 
if random errors follow the assumption,
the effective information of series will be completely utilized to training and fitting the model, thus the results of DW test will approach to 2, suggesting there is no autocorrelation.
Akaike information criterion (AIC) can measure the goodness of data fitting, and the model with smallest score is viewed as the most accurate.

As shown in Fig.~\ref{pqchoose}, AR(0) \& MA(3) model performs well in both the two metrics, especially its DW score is equal to 2.008721 showing the rationality of Gaussian assumption and high efficiency.
Thus we finally choose ARIMA model (0,2,3).

\section{Adaptive Resource Allocation}
As diverse URLLC applications is emerging in the 5G and future communication systems, some services with different QoS requirements will inevitably access and request for the same frequency range simultaneously. 
This calls for more reasonable resource allocation scheme, which would not only fully consider the current users' QoS requirements but also have prediction capability towards the future pressure of resources.
We have derived several network-load estimation and prediction schemes above, and in this section we will discuss the allocation strategy driven by different QoS requirements correspondingly.

\subsection{Access Failure Probability}
We derive the access failure probability with $ W $ RBs, $ N $ users, $ K $ repetitions, and $ T $ slots in an access cycle, which will provide prospective bases for resource allocation.

\subsubsection{Adjacent-type K-Repetition}
The number of active users in the $ t $-th slot is represented as $ n_t $, and the number of users beginning their first access in the $ r $th slot is denoted as $ \phi_r $.
The vector $ \boldsymbol \phi $ denotes $ ({\phi_1}, {\phi_2} , \ldots, {\phi_{T-K+1}})^T $ .
Firstly we target at a randomly chosen user, and at the start of access cycle, it will randomly begin its first attempt at the first $T-K+1 $ slots.
The access failure probability of this target user who begin at the $ r $th slot($ 1 \leqslant r \leqslant { T-K+1} $) can be formulated by:
\begin{equation}
	Pr\left\{\mathrm {tar}\right\}=\prod_{t=r}^{r+K-1}1-\dfrac{\displaystyle\binom{W}{1}{(W-1)}^{n_t-1}}{{W^{n_t}}}\,.\label{target}
\end{equation}

We have derived the equation~(\ref{fangcheng}) above, which depicts the 
relationship between $ n_t $ and $ \phi_r $, and here we refer it as $ f(\boldsymbol\phi) $.
For every determined $ \boldsymbol\phi $, we can use $ f(\boldsymbol\phi) $ to derive the exact number $ n_t $ of users in each slot.

In the following equation, we formulate the total access failure probability, in which $ Pr \left \{ \boldsymbol \phi \, | \, N \right \} $ and $ Pr\left\{\mathrm {tar}\right\} $ have been derived in~(\ref{PMF}) and~(\ref{target}) respectively:
\begin{equation}
	\begin{split}
			Pr\left\{\mathrm {total}\right\}=\sum_{\boldsymbol\phi}\left \{Pr \left \{ \boldsymbol \phi \, | \, N \right \}\left \{\sum_r \dfrac{\phi_r}{N} \;Pr\left\{\mathrm {tar}\right\}\right \}\right \}\,. \label{lin_fail}
	\end{split}
\end{equation}

\subsubsection{Arbitrary-type K-Repetition}
The expected number $ e $ of users in every slot is equal to $ \frac{NK}{T} $,
and the total access failure probability can be derived by:
\begin{equation}
	Pr\left\{\mathrm {total}\right\}=\left \{1-\dfrac{\displaystyle\binom{W}{1}{(W-1)}^{e-1}}{{W^{e}}} \right \}^K \,. \label{ren_fail}
\end{equation}

\begin{figure*}  
	\centering  	
	\includegraphics[width=5in]{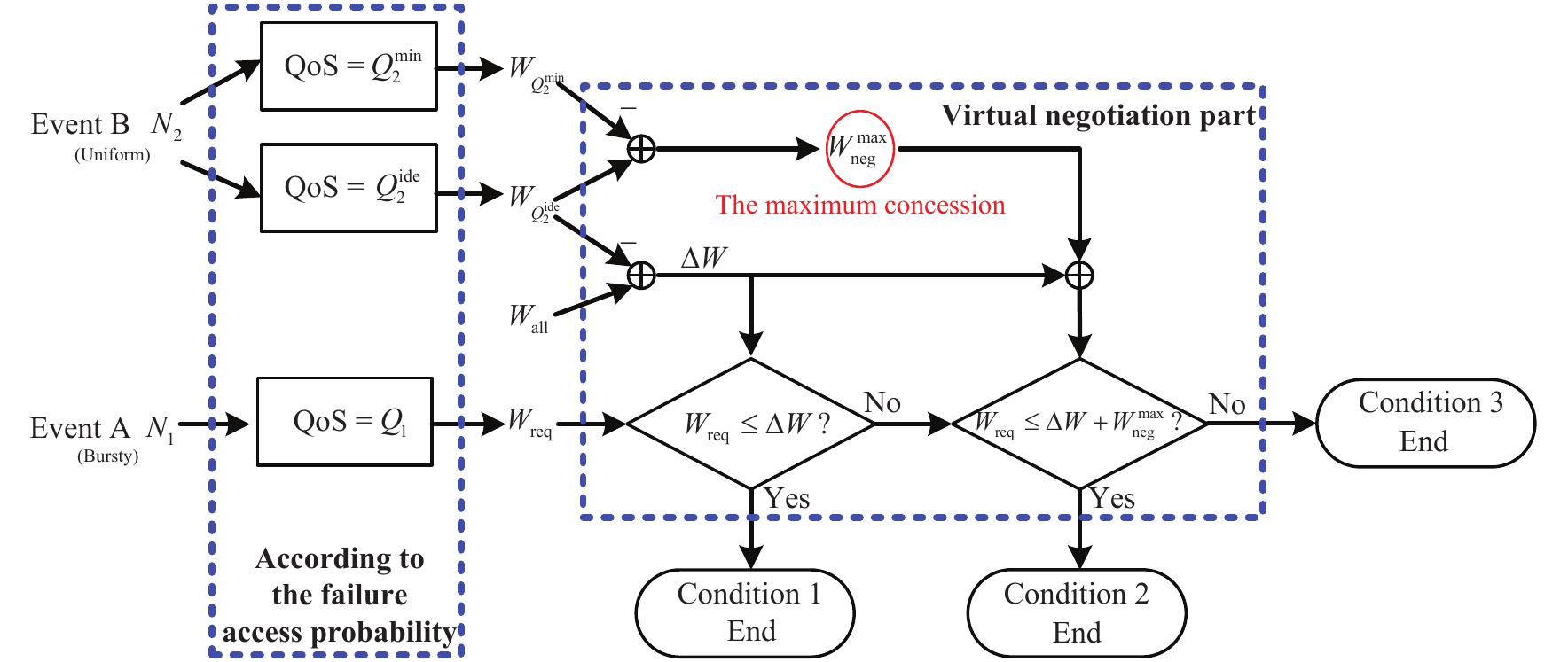}
	\caption{The flow diagram of resource allocation.}  
	\label{allocation} 
\end{figure*} 

\subsection{Allocation Strategy}
\label{allocation_strategy}
We consider two types of services, the one named event A (e.g., following bursty traffic pattern) requires the reliability within 1~ms to meet $ Q_1 $, and the other named event B (e.g., following uniform traffic pattern) requires the reliability within 1~ms to meet $ Q_2^\text{min} $ at least, and its ideal reliability is equal to $ Q_2^\text{ide} $.
The number of users is $ \widehat{N}_1 $ and $ \widehat{N}_2 $ respectively, which can be calculated by the prediction module.
Without loss of generality, $ Q_1 > Q_2^\text{min} $.
Thus the priority of event A is higher than event B.
The whole number of available RBs is denoted as $ W_\text{all} $.
If available RBs are sufficient, we can allocate RBs independently according to each QoS.
However, we also need to discuss resource negotiation considering the conditions when available RBs are inadequate or the system is under high-loads. Moreover, in order to methodically deal with the sudden access, we also allocate a small number of RBs to event A even though its users are inactive~\cite{9569425}.

The aim of resource allocation is to achieve two services' QoS as high as possible, and once the number of available RBs cannot support them simultaneously, the negotiation part would spontaneously sacrifice the uniform service's QoS to satisfy bursty service's QoS. Note that the least permitted QoS of event B should be larger than $ Q_2^\text{min} $.
Thus the maximum number of RBs that can be negotiated is equal to $ W_{Q_2^\text{ide}} - W_{Q_2^\text{min}} $, and we denote it as $ W_\text{neg}^\text{max} $.

We illustrate the allocation logic in Fig.~\ref{allocation}.
For event A, we firstly calculate the required number $ W_\text{req} $ of RBs to meet $ Q_1 $ according to~(\ref{lin_fail}) or~(\ref{ren_fail}).
The realistic condition is divided according to the following rule: 

\begin{itemize}
	\item Condition 1: $ W_\text{req}\,\leqslant\, W_\text{all}- W_{Q_2^\text{ide}}$;
	\item Condition 2: $ W_\text{all}- W_{Q_2^\text{ide}} \,\textless\, W_\text{req} \,\leqslant\, W_\text{all}- W_{Q_2^\text{ide}}+ W_\text{neg}^\text{max} $;
	\item Condition 3: $ W_\text{all}- W_{Q_2^\text{ide}}+ W_\text{neg}^\text{max} \,\textless\, W_\text{req} $.
\end{itemize}

The final number of allocated resource for each service is denoted as $ W_1 $ and $ W_2 $, and the expected QoSs are denoted as $ \widehat{Q}_1 $ and  $ \widehat{Q}_2 $.
We list these parameters under different conditions in Table~\ref{condition}.
In particular, the system will comes to outage in Condition 3, which cannot support the permitted least QoS of each service simultaneously.
One possible result is to meet $ Q_2^\text{min} $ preferentially and allocate the remaining RBs to event A.
\begin{table}[t]
	\vspace{-0.3cm}   
	\centering
	\caption{The advised allocation strategy and expected QoS}   
	\begin{tabular}{m{0.85cm}<{\centering}m{2.45cm}<{\centering}m{1.7cm}<{\centering}m{0.7cm}<{\centering}m{1cm}}    
		\toprule    
		Conditions & $ W_1 $&$ W_2 $&$ \widehat{Q}_1 $&$ \widehat{Q}_2 $\\    
		\midrule   
		Cond.1 & $ W_\text{req}  $& $ W_{Q_2^\text{ide}} $ &$ Q_1 $&$ Q_2^\text{ide} $\\
		\hline   
		Cond.2 & $ W_\text{req}  $& $ W_\text{all}-W_\text{req}  $&$ Q_1 $&$ \geqslant Q_2^\text{min} $\\   
		\hline 
		Cond.3 & $ W_\text{all}- W_{Q_2^\text{ide}}+ W_\text{neg}^\text{max}  $&$
		 W_{Q_2^\text{ide}}- W_\text{neg}^\text{max}  $& $\leqslant Q_1 $ & $ Q_2^\text{min}  $\\    
		\bottomrule   
	\end{tabular}  \label{condition}
\end{table}

\section{Simulation Evaluations}
\subsection{Load Estimation}
\textbf{Baselines I}: \textit{Minimum Square Error for Mean Values of Access States Estimation (MSEM)}

For comparative analysis, we briefly describe another scheme based on \cite{8880524} (for mMTC), and here we modify it to fit the model of URLLC in this paper.
Firstly, a parameter $ \theta_{r}^{e} $ is used to denote the total number of users who access in $ e $ consecutive slots $ (1 <e \leq K) $ starting from $ r $th slot, i.e., $ \theta_{r}^{e}=n_{r}+n_{r+1}+\cdots+n_{r+e-1} $. 
The total number of users in 1ms can be derived by:
\begin{equation}
	\widehat{N}_{\mathrm{tr}}=\frac{1}{K^{2}}\left(\sum\limits_{e=1}^{K-1}\theta_{1}^{e}+ \sum\limits_{r=1}^{T-K+1}\theta_{r}^{K}+\sum\limits_{r=T-K+2}^{T}\theta_{r}^{T-r+1}\right)\,.
\end{equation}

Next, we use $ (A,B,C) $ to describe the total access states of RBs in $ e $ consecutive slots, which the expectations $ \overline{A} $, $ \overline{B} $ and $ \overline{C} $ can be formulated as follows:
\begin{equation}
	\begin{split}
			&\overline{A}=N\left (1-\dfrac{1}{eW}\right )^{N-1} ;\\
			&\overline{B}=eW-\overline{A}-\overline{C};\\
			&\overline{C}=eW\left (1-\dfrac{1}{eW}\right )^{N},
	\end{split}
\end{equation}
where $N $ denotes the test  number of total users for hypothesis in $ e $ slots.
Then,  the BS can reasonably obtain an estimate $ \theta_{r}^{e} $ by minimizing the Euclidian distance between the observed access states of RBs and the theoretical means:
\begin{equation}
	\theta_{r}^{e} =\arg \min \limits_{N} \left[{(A-	\overline{A})^2+(B-	\overline{B})^2+(C-	\overline{C})^2} \right]\,.
\end{equation}

\textbf{Baselines II}: \textit{Idle Resources Counting Estimation (ISCE)}

In~\cite{7072487} Oh \textit{et al}. proposed estimation scheme based on idle resources which can be formulated as follows:
\begin{equation}
	n_t= \dfrac{	\displaystyle\mathrm{log}\frac{C_i}{W}}{	\displaystyle\mathrm{log}\frac{W-1}{W}}\,,
\end{equation}
in which $ n_t $ denotes the number of active users in the $ t $-th slot.
According to the system model described in this paper, we know that $ N_{\mathrm {tr}}$ users with $ K $ replicas contribute to the sum of $n_{t}$, thus we can obtain:
\begin{equation}
	\widehat{{N}} _{\mathrm{tr}}= \dfrac{1}{K}\sum_{t=1}^{T} n_t\,.
\end{equation}

Note that MSEM is suitable to adjacent-occupation $ K $-repetition, and ISCE is suitable to both of two types.
In this subsection, we verify the estimation performances with the results from Monte-Carlo simulations.
For adjacent-occupation $ K $-repetition, the SS-ML-LS, MS-MLI, and baseline I (MSEM) are employed; for arbitrary-occupation $ K $-repetition, the MS-MLD and baseline II (ISCE) are employed.
Furthermore, in order to accelerate our proposed schemes which significantly utilize Markov model, we generate a state table in advance which can support the quick search of transition probabilities. Thus for every hypothetical $ N $ in ML, we only need to query the transition probability with $ N $-steps with this table, rather than calculate $ \mathbf S $ and $ \mathbf P $ repeatedly.

The number range of users is from 8 to 18, and we suppose that these users can be served as sufficient resources considering URLLC applications' high QoS requirements.
Firstly, we simulate a group of users' choices with MATLAB and count the number of success, collision, and idle RBs respectively. Note that this step has no correlation to certain traffic patterns, and the resource states are only determined by the number of users and available resources.
Then all of the estimation schemes will work based on the input $ ({\mathbf A}, {\mathbf B},  {\mathbf C}) $.

\begin{figure}  
	\centering  
	\includegraphics[width=9cm]{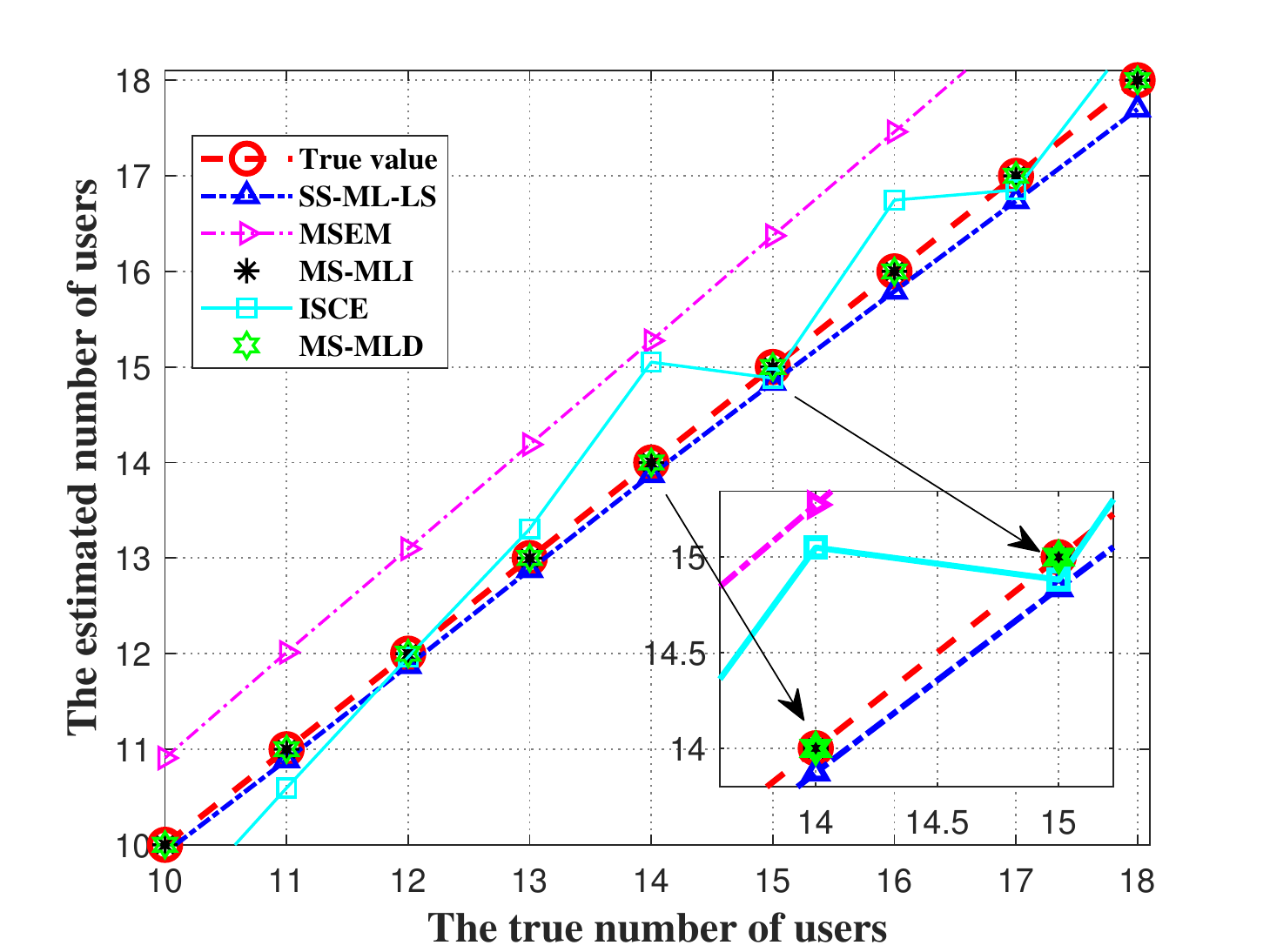}\\  
	\caption{The estimation performances derived from our proposed schemes and baselines}  
	\label{estimation}  
	%\vspace{-0.46cm}
\end{figure}

Fig.~\ref{estimation} depicts the accuracy performances of estimation schemes compared with the true values.
The most accurate estimation schemes are MS-MLI and MS-MLD with almost no bias, which both consider the overall resource states of a complete access cycle simultaneously and thus avoid introducing errors again caused by quadratic estimation.
However, the huge state space of complete access cycle will significantly increase the time complexity searching for the optimal solution, and SS-ML-LS can well offset this problem which estimates in every slots separately.
The estimation accuracy of SS-ML-LS is second only to MS-MLI and MS-MLD.
The baseline MSEM always has large error, while the baseline ISCE cannot perform stably.
In conclusion, if accuracy is a more important factor for estimation, we advice to adopt MS-MLI and MS-MLD; if operation time is more important, we advice to adopt SS-ML-LS.
In the following simulations, we employ the latter.

\subsection{Load Prediction}
\textbf{Baseline}: Moving average with sliding window (MASW)

The $ t $-th expected value is equal to the average of past $ w $ observations, which can be formulated as: 
\begin{equation}
		\widetilde {n}_t=\sum_{i=t-w}^{t-1}n_i\,,
\end{equation}
in which $ n_i ~(t-w \leqslant i \leqslant t-1 ) $ denotes the $ i $th observation selected by the sliding window whose length is equal to $ w $.

\begin{figure}  
	\centering  
	\includegraphics[width=9cm]{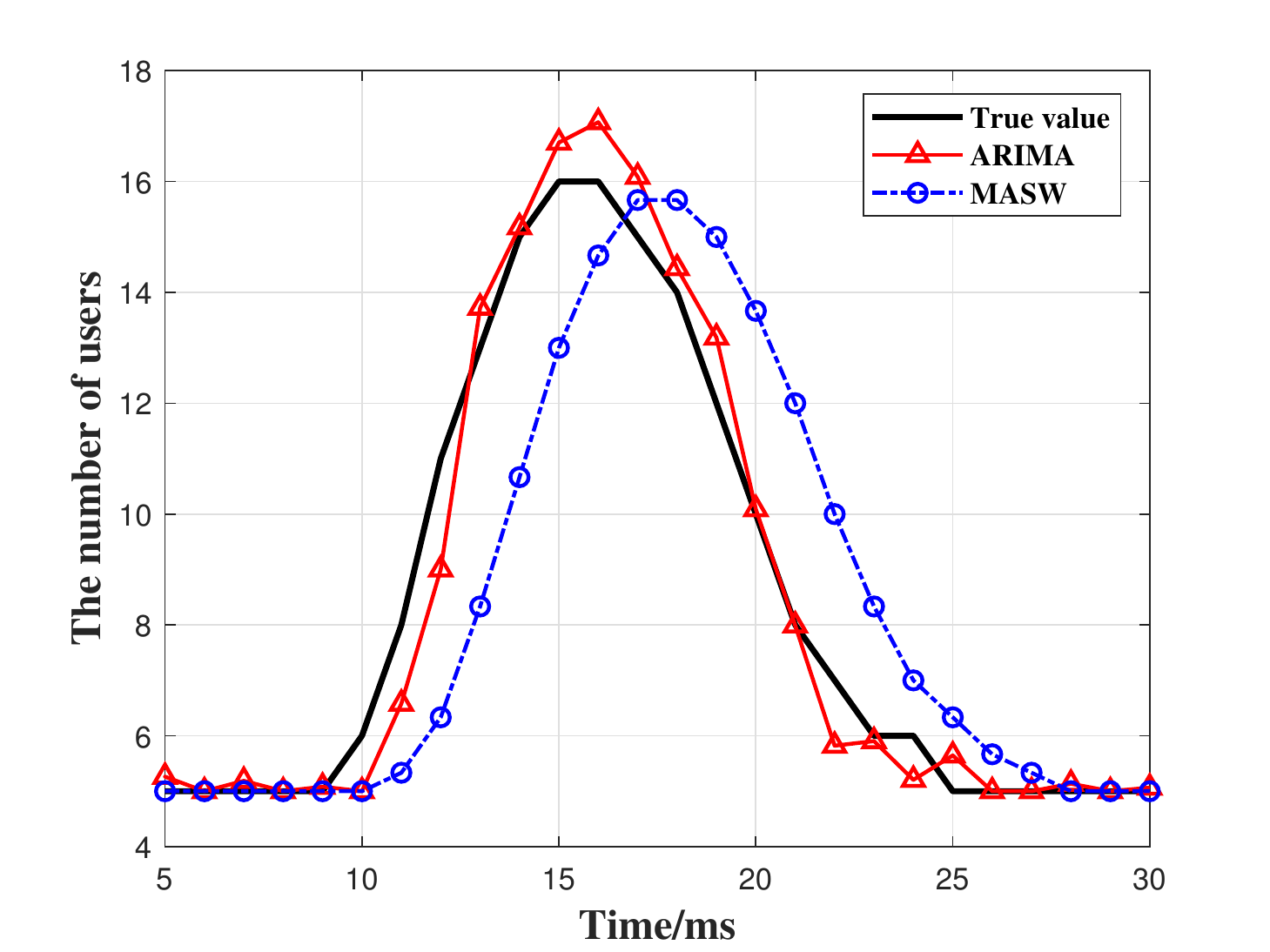}\\  
	\caption{The prediction performances derived from ARIMA and baseline (MASW).}  
	\label{prediction}  
	%\vspace{-0.46cm}
\end{figure}

In this section we employ ARIMA model (0,2,3) and MASW to achieve single-step prediction.
Due to the BS typically has no knowledge on the accurate number of users, the estimated value will be added into the history pool as the realistic observation.
As shown in Fig.~\ref{prediction}, the uniform event lasts for the whole simulation period, and a bursty event happens over 10 $ \sim $ 25~ms which is generated by $ \mathcal{B}(80,15) $. The average prediction error by ARIMA is equal to 6.8\%, and that by MASW is equal to 21.9\%. From the perspective of global fluctuation, ARIMA can sense and response to the rise and fall tendency of observations in time, while there are always lags between the predictions derived by MASW and realities. 

\subsection{Access Failure Probability}
\begin{figure}  
	\centering  
	\subfigure[]{
		\includegraphics[width=9cm]{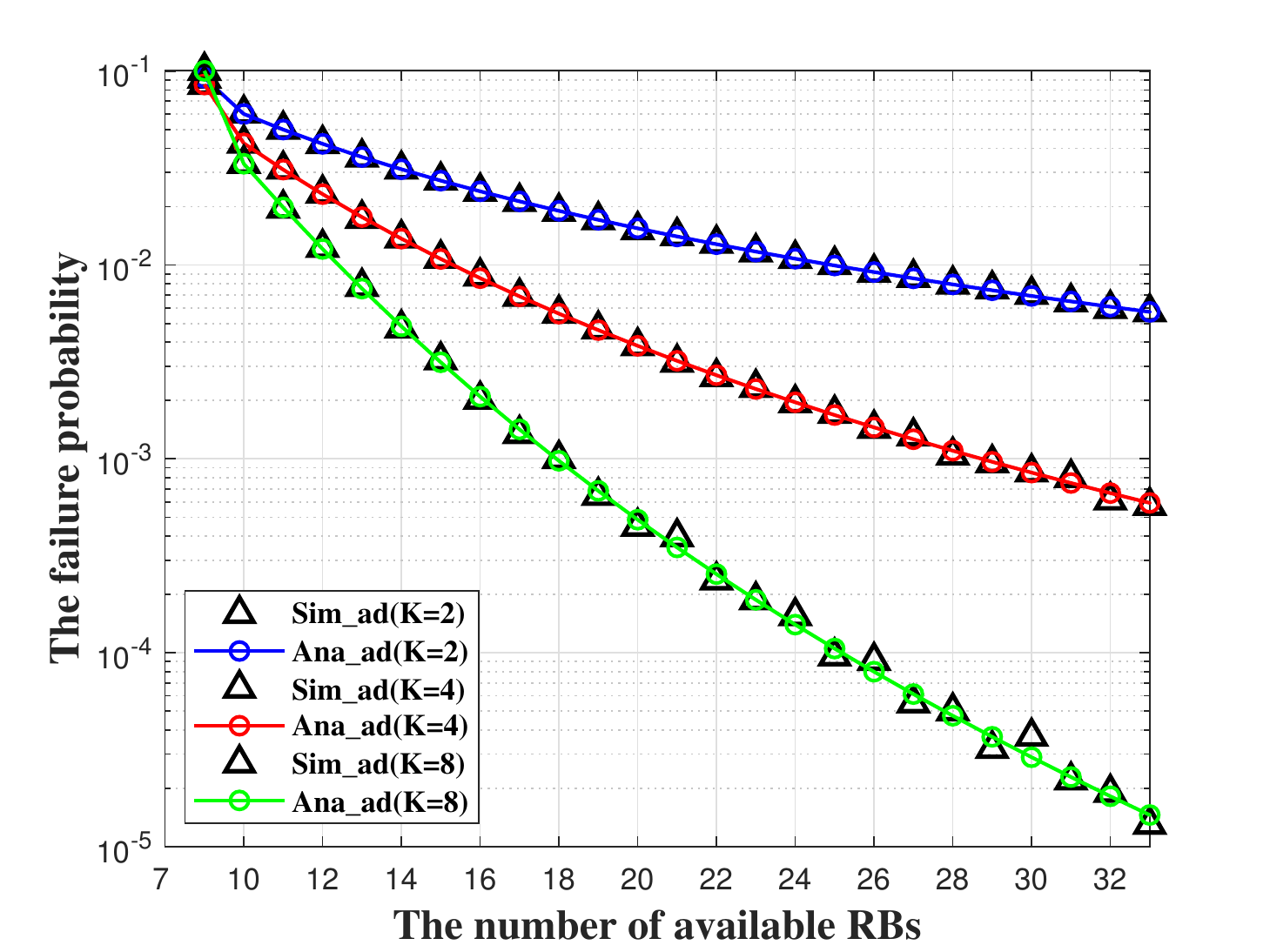} 
		\label{fail_lin}
}
	\subfigure[]{
	\includegraphics[width=9cm]{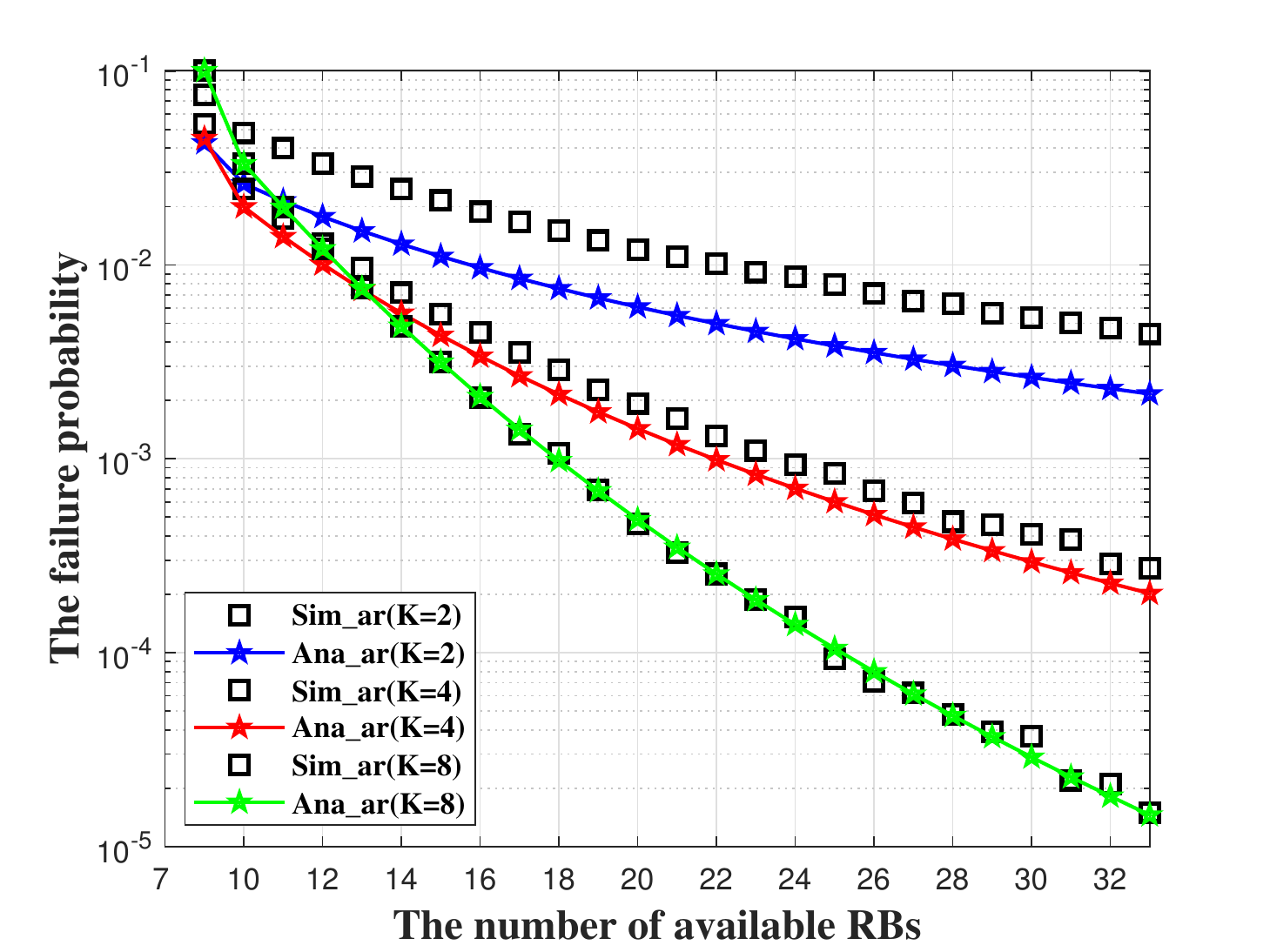}  
	\label{fail_ren}
}
	\caption{The number of available RBs versus the failure probabilities. Here we compare the analytical results with simulation results and also discuss several variables including $ K $-repetition types and repetition times. (a) discusses the adjacent-occupation $ K $-repetition with $ K=2,4,8 $, and (b) discusses the arbitrary-occupation $ K $-repetition with $ K=2,4,8 $.}  
	\label{failureprob}  
	%\vspace{-0.46cm}
\end{figure}

In Fig.~\ref{failureprob}, we compare the analytical results of access failure probability with simulation results and also discuss performances of two types of $ K $-repetition schemes with different $ K $ values. The number of active users is set as 10, and the range number of  available RBs is from 7 to 33.
 
Firstly for adjacent-occupation $ K $-repetition calculated by Eq.~(\ref{lin_fail}), analytical results (i.e., Ana\_ad) are very close to simulation results (i.e., Sim\_ad)  with errors ranging in 0.52\% \~{} 0.74\%, as shown in Fig.~\ref{fail_lin}. 
For arbitrary-occupation $ K $-repetition calculated by Eq.~(\ref{ren_fail}),
as repetition times are closer to the total number of slots in an cycle, the analytical results (i.e., Ana\_ar) will be more accurate, as shown in Fig.~\ref{fail_ren}. 

Moreover, with the same $ K $ and repetition times, arbitrary-occupation $ K $-repetition can achieve better access performances compared with adjacent-occupation $ K $-repetition. This is because arbitrary-type permits users to access with higher degrees of freedom and thus reduces collisions. When $ K=8 $, i.e., a user will utilize all the available slots of a an access cycle, the arbitrary-occupation $ K $-repetition is equivalent to adjacent-type.

In the $ K $-repetition access scheme, on the one hand, the replicas can enhance success probabilities, while on the other hand, excessive repetitions will also lead to frequent collisions and decline success probabilities instead.
Considering the resources allocated to URLLC applications are sufficient,
the users can access with relatively high $ K $ values without intense collisions.
In  Fig.~\ref{failureprob} we can notice that with the same number of available RBs, either in arbitrary-type or adjacent-type, the larger the repetition times $ K $ is, the better the access performances are.

\subsection{Negotiation in Adaptive Resource Allocation}
\begin{figure}  
	\centering  
	\includegraphics[width=9cm]{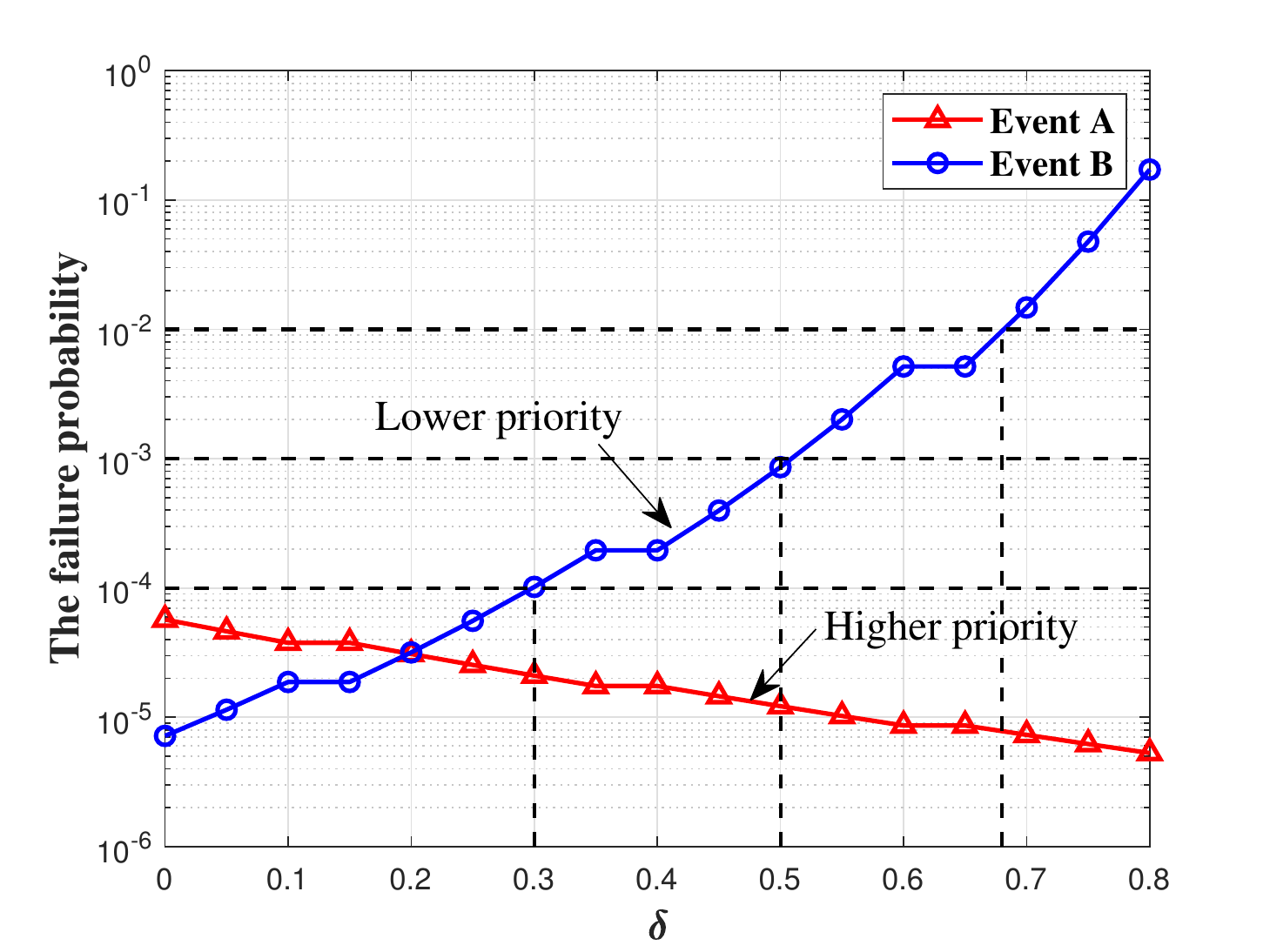}\\  
	\caption{The parameter $ \delta $ versus failure probabilities of event A and B. The negotiated number of resources allocated to event A is equal to $ W_A*(1-\delta)$, while that to event B is equal to $ W_B+W_A*\delta $. Here if reliability requirements of event A and B are 99\% and 99.999\%, we can set $ \delta $ as 0.55.}  
	\label{negotiation}  
	%\vspace{-0.46cm}
\end{figure}

We illustrate the negotiation part of allocation scheme via Fig.~\ref{negotiation}.
Here we consider two applications with different QoS requirements, i.e., event A and event B. Without loss of generality, we assume the event B has higher priority than event A.
The $ W_A $  and $ W_B $ denote the resources allocated to event A and event B originally.
If $ W_B $ cannot satisfy event B's QoS, the negotiation part will be switched on.
For brief description, the parameter $ \delta $ is used to control negotiation part. In particular, the negotiated number of resources allocated to event A and event B is $ W_A*(1-\delta)$ and $ W_B+W_A*\delta $.
Thus in the actual operation, after we get the predicted number of users for the nest cycle, we can calculate the corresponding failure probabilities versus $ \delta $, then find the most appropriate $ \delta $ considering different QoS limitations.
For instance, in the Fig.~\ref{negotiation}, if the minimum reliability requirements of event A and event B are 99\% and 99.999\% respectively, $ \delta $ should be 0.55;
If the minimum reliability requirements of event A and event B are 99.99\% and 99.999\% respectively, the total available RBs are insufficient thus it comes to outage.  

\subsection{Integrated Simulation}
\begin{figure*}  
	\centering  
	\subfigure[Bursty event with Setup 1]{
		\includegraphics[width=8.5cm]{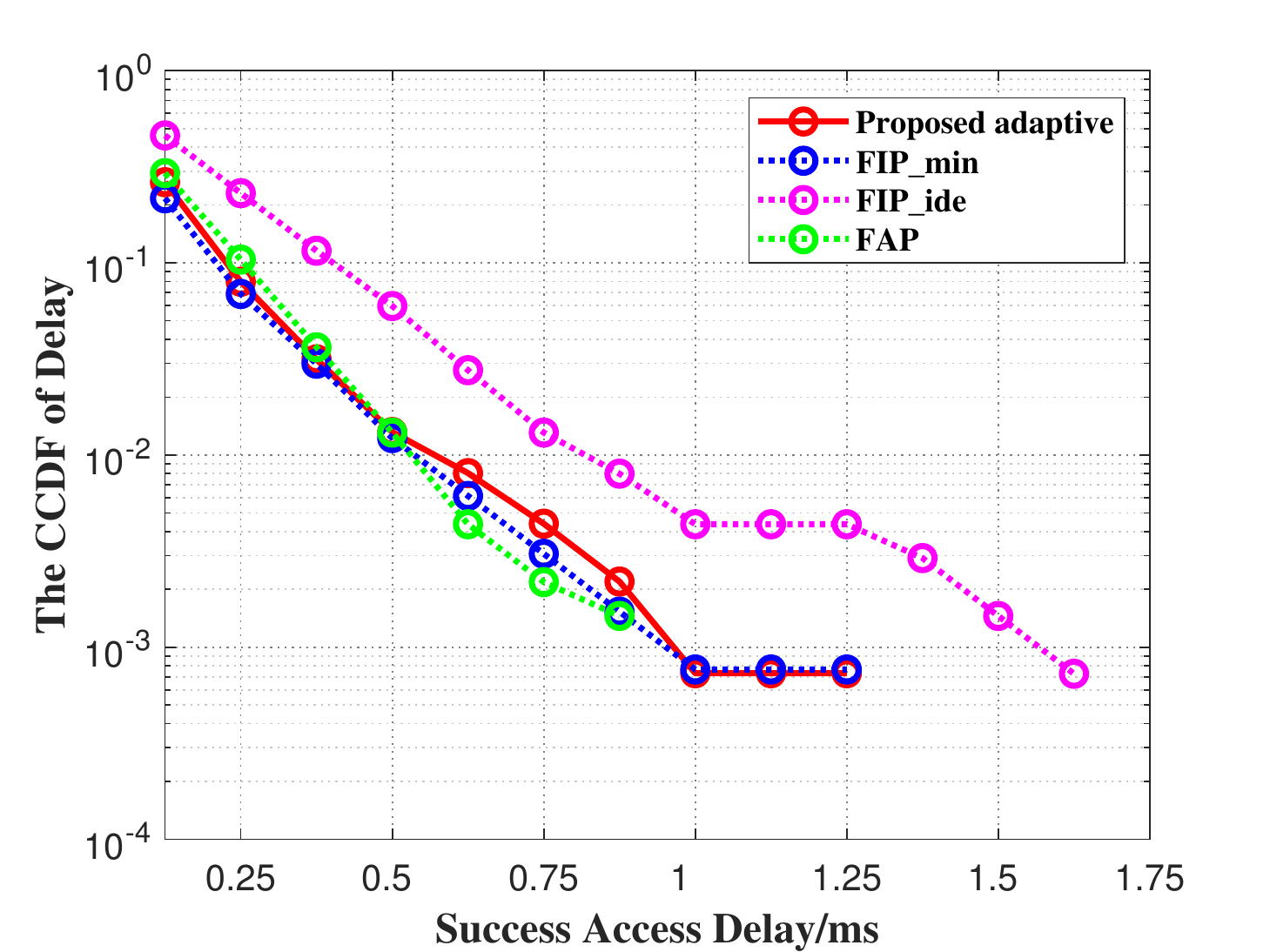} 
		\label{inter1_bursty}
	}
	\hspace{-0.5cm}
	\subfigure[Uniform event with Setup 1]{
		\includegraphics[width=8.5cm]{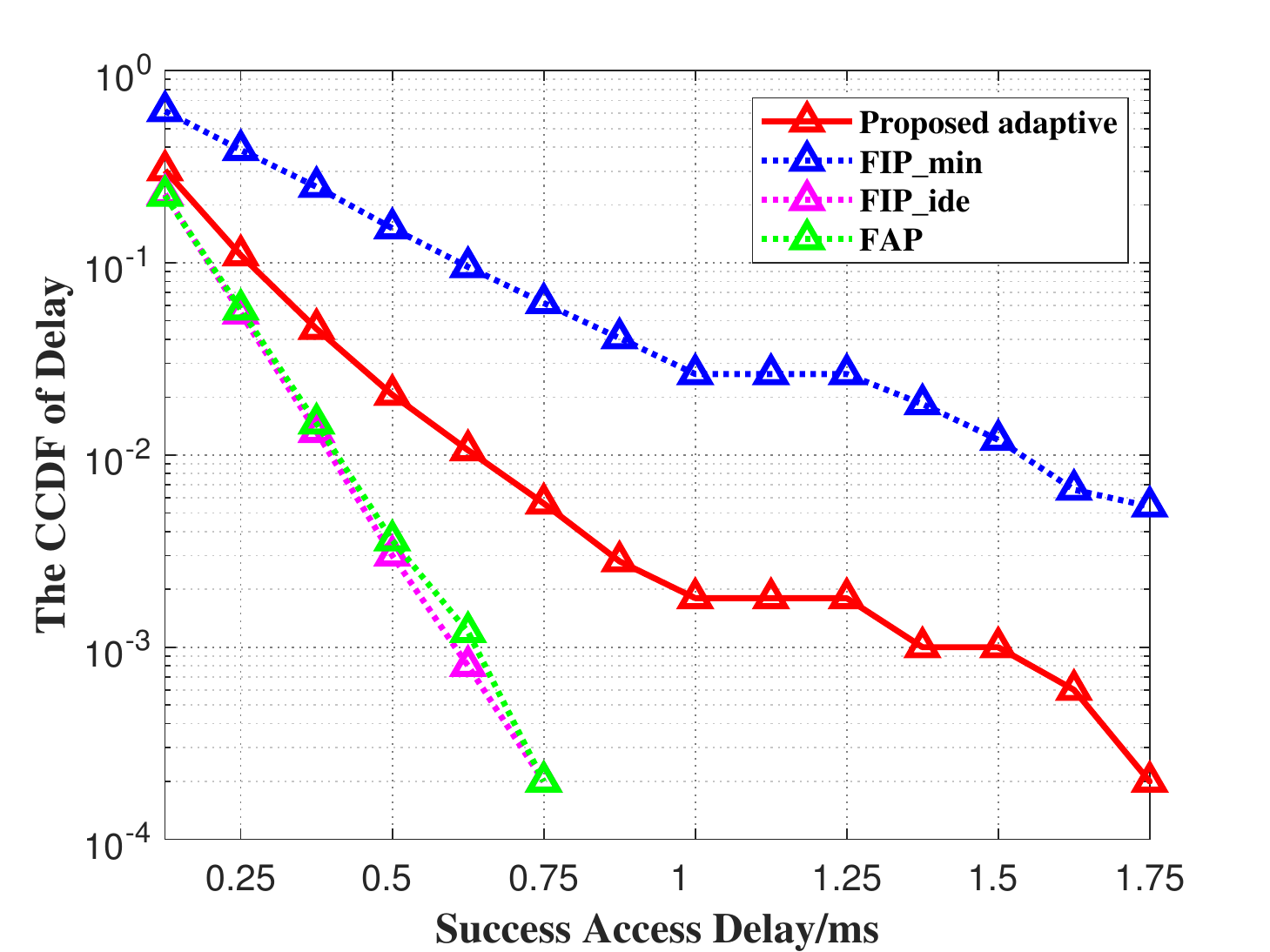}  
		\label{inter1_uniform}
	}
	%\hspace{-0.5cm}
	
	\subfigure[Bursty event with Setup 2]{
		\includegraphics[width=8.5cm]{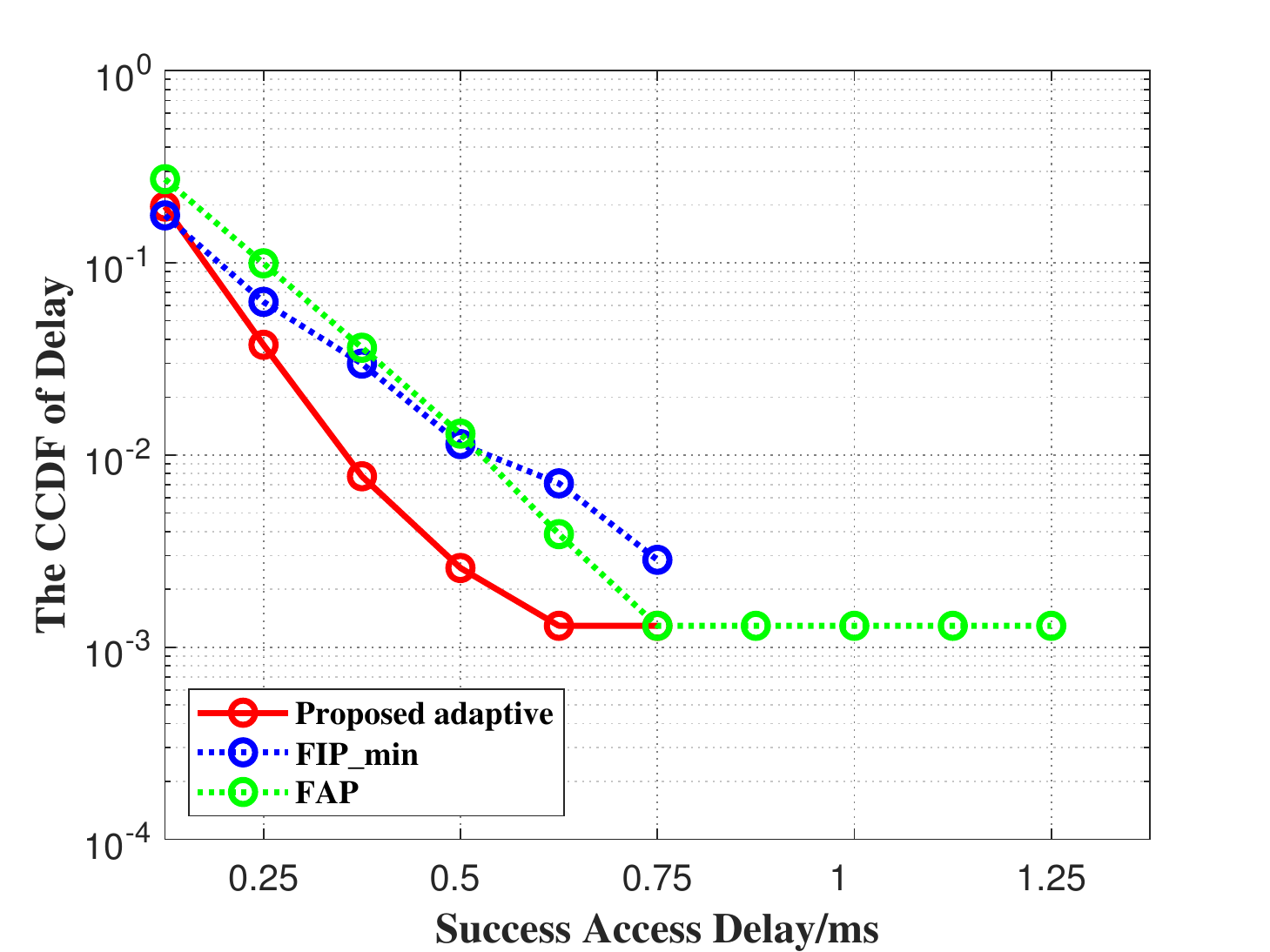} 
		\label{inter2_bursty}
	}
	\hspace{-0.5cm}
	\subfigure[Uniform event with Setup 2]{
		\includegraphics[width=8.5cm]{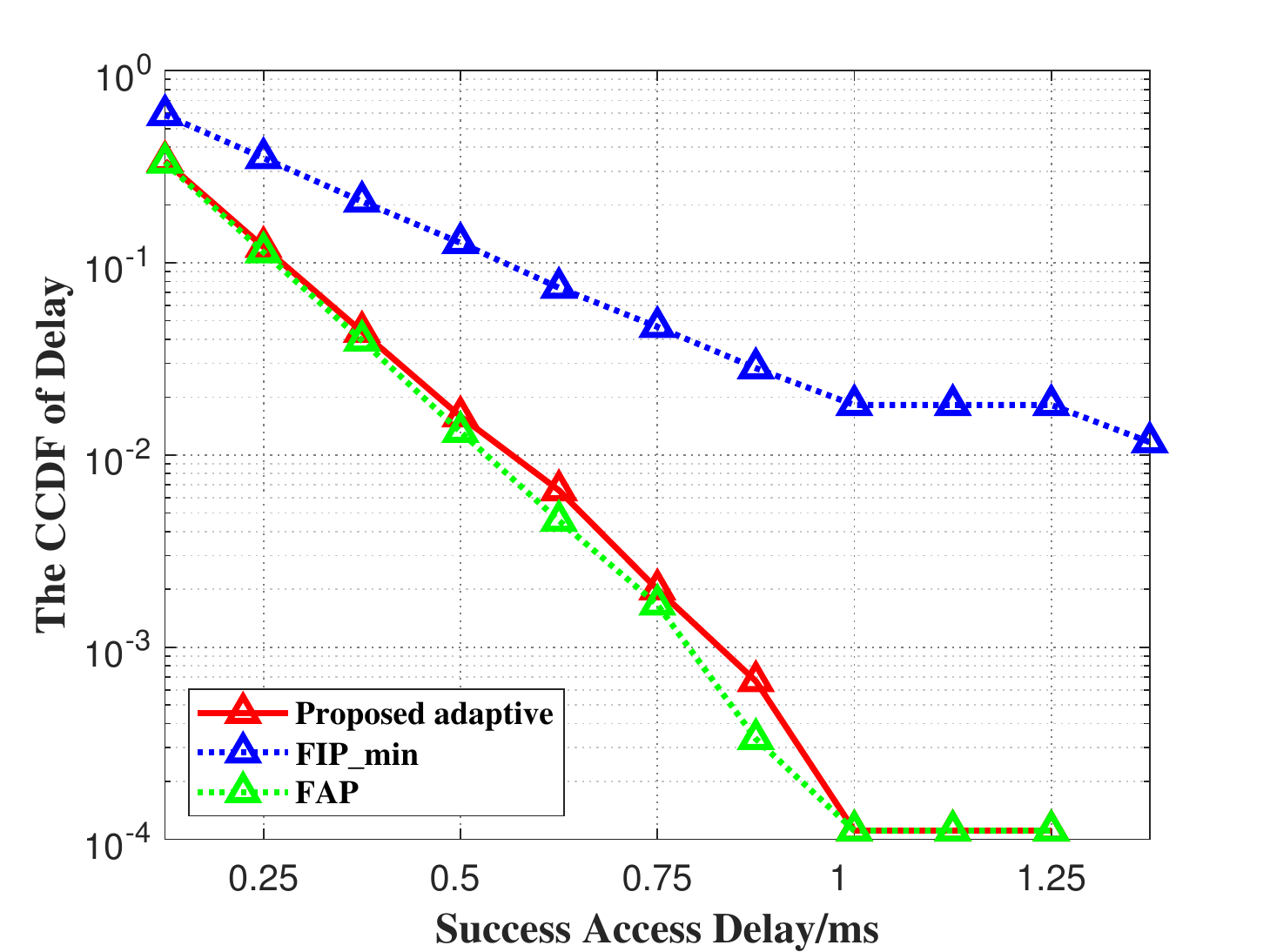}  
		\label{inter2_uniform}
	}
	\caption{The success access delay of our proposed adaptive allocation scheme compared with baseline $ \mathrm{FIP\_ _{min}} $,  $ \mathrm{FIP\_ _{ide}} $, and FAP. (a) and (b) are derived with parameter Setup 1, while (c) and (d) are derived with parameter Setup 2. Note that $ \mathrm{FIP\_ _{min}} $ means the fixed RBs of uniform users are determined according to their minimum QoS, and $ \mathrm{FIP\_ _{ide}} $ means the ideal QoS.}
	\label{integrated}  
	%\vspace{-0.46cm}
\end{figure*} 

In this part we will simulate to verify the performances of complete framework proposed in this paper.
We set the following methods for comparison:
1) Fairy proportion allocation (FAP): The requirements of various services will be considered fairly rather than according to priorities. The RBs allocated to each service is determined according to the ratio of the number of this service's users to the total number of users.
2) Fixed priority allocation (FIP): The event with lower priority will be allocated with fixed RBs according to its idle QoS ($ \mathrm{FIP\_ _{ide}} $) or minimum QoS ($ \mathrm{FIP\_ _{min}} $), thus the remaining RBs can be reserved for the bursty event with higher priority.
Here we suppose the baselines FAP and FIP have the same knowledge of predicted users as our proposed framework.

Specifically, the simulation scenario is composed by two types of services, i.e., uniform and bursty event. We have introduced them in Section~\ref{traffic_patterns} and Section~\ref{allocation_strategy}. 
Due to the BS is able to designate RBs to them separately, the access states of each service can be observed independently.
The average intensity of beta distribution ranges in 0.5 $ \sim $ 6 users every 1~ms typically~\cite{7447749}. Here we set two groups of arrival intensities, as shown in Table~\ref{parameters}. 
For clear comparison, we employ the  empirical complementary cumulative distribution functions (CCDF) of delay as an evaluation index. Note that there are two idle slots between two cycles for the BS to broadcast, thus the CCDF of 1 $ \sim $ 1.25ms remain unchanged.

\begin{table}[]
	\vspace{-0.3cm}
	\centering
	\caption{Integrated Simulation Parameters}
	\label{parameters}
	%\rowcolors{2}{gray!25}{white}
	\begin{tabular}{m{2.5cm}<{\centering}| m{2cm} <{\centering}m{2cm}<{\centering}}
		\hline  % 顶部线
		Parameters & Setup 1 & Setup 2\\
		\hline Uniform intensity & 10 users & 18 users\\
		\hline Bursty intensity & 50 users in 10 cycles & 25 users in 10 cycles\\
		\hline $ Q_{\text u}^{\text {min}} $ & \multicolumn{2}{c}{99\% reliability within 1~ms} \\
		\hline $ Q_{\text u}^{\text {ide}} $ & \multicolumn{2}{c}{\multirow{2}{*}{99.999\% reliability within 1~ms} }\\
		\cline{1-1}
		$ Q_{\text b} $ & \multicolumn{2}{c}{}\\
		\hline $ W_{\text{all}}$ & \multicolumn{2}{c}{48 RBs} \\
		\hline $ T $ & \multicolumn{2}{c}{8 slots in 1~ms} \\
		\hline  $ K $ & \multicolumn{2}{c}{8 repetitions}\\
		\hline
	\end{tabular}
\end{table}

The simulation results are presented in Fig.~\ref{integrated}.
In Fig.~\ref{inter1_bursty} and Fig.~\ref{inter1_uniform}, $ \mathrm{FIP\_ _{ide}} $ scheme achieves the best QoS for uniform users at cost of serious loss of bursty users' QoS.
On the contrary, $ \mathrm{FIP\_ _{min}} $ scheme enhances bursty users' QoS compared with $ \mathrm{FIP\_ _{ide}} $, however, its uniform QoS is much lower than other schemes.
Our proposed adaptive scheme can achieve 99.999\% reliability for bursty users within 1.375~ms, which is similar to $ \mathrm{FIP\_ _{min}} $ scheme. At the same time, its uniform QoS is also higher than $ Q_{\text u}^{\text {min}} $. Because when there is no bursty event, the uniform users will have the right to flexibly utilize more RBs than that in $ \mathrm{FIP\_ _{min}} $ scheme.
For FAP scheme, it seems to achieve perfect performances both in bursty event and uniform event. However, this is because the ratio $ \frac{N_\text{bur}}{N_\text{bur}+N_\text{uni}}$  under parameter group 1 coincidentally makes reasonable divisions of total resources.

In Fig.~\ref{inter2_bursty} and Fig.~\ref{inter2_uniform}, we won't discuss $ \mathrm{FIP\_ _{ide}} $ in that 18 uniform users calling for $ Q_{\text u}^{\text {ide}} $ will occupy almost all the RBs which causes bursty users' QoS unbearable.
Due to the number of uniform users is much larger than bursty users under parameter group 2, most of the available RBs are allocated to the uniform event in FAP scheme. The more unbalanced the ratio is, the more evident this tendency is.
Under this condition, our proposed scheme is still better than $ \mathrm{FIP\_ _{min}} $ scheme assured by flexible allocation.

\section{Conclusions}
In this paper, we considered $ K $-repetition Grant-Free access in URLLC services and proposed a multi-tier-driven computing framework to assure different QoS requirements. 
In the first tier we designed three network-load estimation schemes, which can estimate the number of current active users based on the resource states (success, collision, and idle).
In the second tier we formulated adaptive resource allocation scheme. We employed ARIMA model to predict loads for the next cycle firstly. Moreover, we also derived analytical formulations of access failure probability within 1~ms for $ K $-repetition access.
Then the allocation scheme would calculate the reasonable RBs driven by different QoS requirements.
Our simulation results showed that MS-MLD and MS-MLI were the most accurate schemes with almost no error, the ARIMA model could achieve accurate and timely predictions with 6.8\% relative error, the analytical formulations had relative errors lower than 1\% compared with simulation results.
Finally, in the integrated simulation, we verified the flexibility and rationality of adaptive resource allocation facing the different access intensities and QoS requirements compared with other baselines.

\bibliographystyle{IEEEtran}%规定参考文献的样式
\bibliography{reference}  %参考文献库的名字Ref
\end{document}